\documentclass[twocolumn,journal]{IEEEtran}
\usepackage{ifpdf}
\usepackage{url}
\usepackage{amsmath}
\usepackage{subfigure}
\usepackage{comment}
\usepackage{color}
\usepackage{enumerate}
\usepackage{cite}
\usepackage{amssymb}
\usepackage{bbm}
\usepackage{mleftright}
\usepackage{booktabs}
\usepackage{diagbox}
\usepackage{hhline}
\usepackage{balance}
\usepackage[dvipsnames]{xcolor}

\usepackage{amsthm}

\ifCLASSINFOpdf
  \usepackage[pdftex]{graphicx}
  \DeclareGraphicsExtensions{.pdf,.jpg,.png,.eps,.png}
\else
\fi

\begin{document}

\title{Handling Spontaneous Traffic Variations in 5G+ via Offloading onto mmWave-Capable UAV `Bridges'}



\author{Nikita~Tafintsev, Dmitri~Moltchanov, Sergey~Andreev, Shu-ping~Yeh,\\
Nageen~Himayat, Yevgeni~Koucheryavy, and Mikko~Valkama
\vspace{-4mm}
\thanks{Copyright \textcopyright 2020 IEEE. Personal use of this material is permitted. However, permission to use this material for any other purposes must be obtained from the IEEE by sending a request to pubs-permissions@ieee.org. }
\thanks{N. Tafintsev, D. Moltchanov, S. Andreev, Y. Koucheryavy, and M. Valkama are with Tampere University, Finland. Email:~{firstname.lastname}@tuni.fi}
\thanks{S.-p. Yeh and N. Himayat are with Intel Corporation, Santa Clara, CA, USA. Email:~{firstname.lastname}@intel.com}
\thanks{This work was supported by Intel Corporation, project 5G-FORCE, and Academy of Finland (project RADIANT).}
\vspace{-4mm}
}

\maketitle

\begin{abstract}

Unmanned aerial vehicles (UAVs) are increasingly employed for numerous public and civil applications, such as goods delivery, medicine, surveillance, and telecommunications. For the latter, UAVs with onboard communication equipment may help temporarily offload traffic onto the neighboring cells in fifth-generation networks and beyond (5G+). In this paper, we propose and evaluate the use of UAVs traveling over the area of interest to relieve congestion in 5G+ systems under spontaneous traffic fluctuations. To this end, we assess two inherently different offloading schemes, named routed and controlled UAV `bridging'. Using the tools of renewal theory and stochastic geometry, we analytically characterize these schemes in terms of the fraction of traffic demand that can be offloaded onto the UAV `bridge' as our parameter of interest. This framework accounts for the unique features of millimeter-wave (mmWave) radio propagation and city deployment types with potential line-of-sight (LoS) link blockage by buildings. We also introduce enhancements to the proposed schemes that significantly improve the offloading gains. Our findings offer evidence that the UAV `bridges' may be used for efficient traffic offloading in various urban scenarios.
\end{abstract}

\section{Introduction}\label{sect:intro}

\subsection{Background and Motivation}


The Third Generation Partnership Project (3GPP) is currently ratifying a new fifth-generation (5G) radio interface referred to as New Radio (NR) \cite{3gpp38300}. By now, 3GPP completed the standalone 5G NR specifications, which enable detached NR-based deployments \cite{3gpp180516}. Full compliance with International Mobile Telecommunications-2020 (IMT-2020) requirements will be part of 3GPP Release 16. While network operators are preparing for the deployment of NR infrastructure, 3GPP and academia are exploring new options and use cases of developing networks.


5G networks are characterized by stringent performance requirements in terms of data rate, latency, and reliability. According to the latest 3GPP NR specifications, millimeter-wave (mmWave) operation is one of the key technologies in 5G \cite{3gpp38802}. Such layouts have tremendous potential for rate and capacity. However, it is impractical to deploy mmWave cells in macro scenarios, due to inability of mmWaves to penetrate walls, buildings, or other obstacles \cite{8386686}, \cite{7999294}. Therefore, for customers to take full advantage of mmWave technology, dense deployments of cells will be required. Target scenarios may include city squares, crowded crossroads, or busy parks. In these use cases, traffic demand is highly dynamic with occasional spikes in the offered load. Therefore, it is difficult to predict expected load levels since user requests are distributed unpredictably across space and time.


One of the ways to serve spatio-temporal traffic demand is to assist the cellular network by offloading excessive traffic. As a possible solution for flexible and dense deployments of the 5G networks, integrated access and backhaul (IAB) technology has recently emerged. Originally, the advantages of IAB architectures were shown in \cite{3gpp171880}. Furthermore, the principle was advanced in \cite{3gpp38874}, which focuses on IAB with physically fixed relays. With the introduction of such systems, unmanned aerial vehicles (UAVs) equipped with the IAB capabilities may facilitate on-demand network densification. By integrating UAVs as IAB nodes, significant enhancements can be made to the coverage, connectivity, or capacity of wireless networks \cite{8690860}, \cite{8641421}. To further leverage data relaying benefits, multi-hop support is presently being discussed by 3GPP (see \cite{3gpp1806814} and \cite{3gpp1806815}). It becomes feasible due to higher directivity of NR communications as this efficiently reduces the interference \cite{kulkarni2018max}. Versatile multi-hop data relaying topology is assumed to be one of the key components in prospective 5G+ systems to connect bandwidth-hungry users with the core network.


Not limited to the telecom sector, UAVs bring advantages in various other fields like transportation, agriculture, and medicine~\cite{gsmauav}. For example, commercial use of UAVs in the delivery industry improves efficiency, lowers costs, and enhances customer experience with potentially life-saving benefits in a variety of scenarios. UAVs effectively solve the expensive last-mile problem by sending supplies across the cities or to remote areas. The utilization of UAVs provides an option for on-demand and same-day delivery as well as the ability to avoid limitations of traditional logistics, such as roadway delays. For instance, partnering with a medical supply company, UPS utilized UAVs for on-demand emergency deliveries. Jointly with a UAV manufacturer, they initiated a medical-sample delivery system for hospitals in North Carolina, USA~\cite{wired}. In April 2019, Google's project ``Wing'' has become the first UAV initiative in the USA to receive governmental approval for goods delivery \cite{bloomberg}. Presently, UAV regulations still do not permit most of the flights over crowds and in urban areas, thus limiting operation. However, the regulations are becoming increasingly flexible for the companies using UAVs.


In addition to their main mission, UAVs conducting delivery can perform supplementary tasks. For example, Amazon acquired a patent for the use of delivery UAVs as flying surveillance cameras for residential buildings \cite{uavimagecreation}. Here, the main task of the UAV is delivery. However, during the shipping process, if the UAV has additional resources, it will be able to carry out the planned monitoring of private property. It is expected that the number of such applications will grow.


Motivated by the integration of UAVs as multi-hop IAB nodes and other emerging use cases such as load balancing and congestion avoidance, this paper studies data offloading schemes that employ UAVs in various 5G+ deployments while concentrating on connectivity time and offloading efficiency.

\subsection{Related Work}


Over the recent years, the research community made an enormous effort in supporting the integration of UAVs into the modern cellular infrastructure. Starting from Rel. 15, 3GPP introduced the corresponding functionality into their wireless systems. In this regard, \cite{3gpp22829} reviews a wide range of usage scenarios and examines the UAV features that may demand additional support. This technical report defines the expected service-level requirements for accommodating various UAV use cases by the 3GPP system. From the wireless communication perspective, the use of radio access capabilities onboard UAVs has attracted interest from the community~\cite{8644097, 8403671, tafintsev2019aerial, 8506622}. Currently, the latest efforts continue to investigate new techniques for further UAV support. This includes work on NR relaying with IAB capabilities \cite{3gpp38874}, which has been discussed in Rel. 15 and now continues in Rel. 16.


The research work on wireless backhauling over UAVs has accelerated recently. In \cite{7572068, 5937283, 8246580, 8643815}, a UAV-enabled relaying system was analyzed where a UAV is acting as a relay node for communication between wireless devices. In~\cite{7932157}, the authors studied the scenarios where UAVs offer computation offloading opportunities to mobile users. Their results confirmed notable energy savings under the introduced optimization method of bit allocation and trajectory planning as compared to mobile implementation. The authors in~\cite{8641421} investigated mobile-enabled UAVs and wireless infrastructure UAVs within the realistic constraints of 5G networks. They showed that UAVs serving as intermediate relay nodes may potentially bring additional gains to the levels of wireless connectivity. Further, due to their mobility, UAVs can optimize the number of hops and improve topology flexibility. The recent findings in 3GPP contributions suggest considerable benefits from the use of single-hop relaying. Going further, multi-hop relaying topologies are considered as one of the key interest areas in future 5G+ systems.


There is a growing number of works related to multi-hop UAV networks. In \cite{8254715}, the authors proposed a backhaul scheme that employs UAVs as an on-demand flying network. Their simulation results argued that the proposed approach demonstrates significant performance gains in terms of the data rate and transfer delay. In \cite{8453022}, the authors considered the multi-hop relaying system where multiple UAVs serve as aerial relays to assist the communications between the ground nodes. They proposed an algorithm for maximizing the end-to-end throughput performance, the efficiency of which was validated with supportive numerical results. In \cite{7510735}, the authors developed a network model for transmitting real-time observed data in a multi-hop relay fashion. They investigated real-time data transmission over a UAV network and evaluated system performance. In \cite{8424236}, the challenges of multi-hop UAV-assisted emergency scenarios were discussed. The authors optimized trajectory and scheduling of UAVs to provision wireless service for ground devices. The authors in \cite{8764452}, by applying the convex approximation techniques, jointly optimized the UAV’s trajectory and transmit power for covert communications. In \cite{kulkarni2018max}, the authors studied mmWave deployments with multiple backhaul hops. They modeled a mesh network in the urban-canyon layout and provided important design guidelines for routing and scheduling strategies. These latest results highlight the potential of multi-hop UAV networks. However, there are only a few studies so far that envision mmWave-capable multi-hop UAV relaying capabilities.

\subsection{Main Contributions}


In this study, we envision that network operators may utilize UAVs for offloading excessive traffic from overloaded NR base stations (BSs). According to this thinking, network operators employ a fleet of UAVs, which are used for connectivity purposes. Alternatively, network operators may have limited operation capabilities over third-party UAVs (e.g., those used for goods delivery) and may establish, for example, dynamic service level agreements (SLAs) with the UAV operators by specifying configuration of the service.

Instead of relying upon complex mesh typologies, we advocate for the use of the so-called mmWave-capable UAV `bridges', where UAV following the same route within the deployment area operate in a multi-hop relay mode by establishing chain-like topologies that connect currently overloaded and underloaded NR BSs. Using the tools of stochastic geometry and renewal processes, we characterize the offloading schemes that will or will not violate the key performance indicators (KPI) of the UAV operator. Taking into account mmWave propagation features and blockage by buildings, we characterize the UAV connectivity process. First, we study a routed UAV `bridging' scheme, where the network operator may affect the choice of routes. Then, we consider a controlled UAV `bridging' scheme, where inter-UAV distances are allowed to be fully managed by a network operator.



The main contributions of this paper are
\begin{itemize}
  \item{We propose a novel approach for offloading the excessive traffic demand by using UAVs traveling over the area of interest. In this approach, multiple consecutive UAVs form a communication `bridge', which connects the overloaded NR BSs with the underloaded ones. It is shown that the proposed approach significantly improves system performance by offloading a considerable fraction of ground user demand.}
  \item{We develop a mathematical methodology, which captures the offloading gain of the proposed approach, including the connectivity properties and the fraction of the requested data that can be offloaded. It is demonstrated that due to mmWave limitations, the UAV connectivity process heavily depends on the distance between the adjacent UAVs and the line-of-sight (LoS) blockage by buildings at the intersections.}
  \item{To confirm the applicability of the developed methodology, we propose several enhancements to  the considered schemes that dramatically improve the offloading factor and counteract the effects of LoS link blockage as well as minimize the number of intersections along the path.}
\end{itemize}


The rest of the paper is organized as follows. First, we introduce the system model and the UAV offloading strategies in Section \ref{sect:system}. Then, the considered offloading schemes are analyzed in Section \ref{sect:framework}. Finally, selected numerical results are provided in Section \ref{sect:numerical}, and conclusions are drawn in Section \ref{sect:conclusions}.

\section{System Model}\label{sect:system}

In this section, we formulate the system model by specifying its components, which include deployment, signal propagation, blockage, antennas, traffic, and UAV dynamics models. Finally, we specify our metrics of interest. The notation used in this paper is provided in Table \ref{table:notation}.

\begin{table}
\renewcommand{\arraystretch}{1}
\caption{Notation used in this work.}
\label{table:notation}
\centering
\begin{tabular}{p{0.15\columnwidth}|p{0.75\columnwidth}}
\hline
\hline
\bfseries Parameter & \bfseries Definition \\
\hline
N & Number of streets in the considered deployment \\
\hline
$d$ & Street width \\
\hline
$d_B$ & Building block length \\
\hline
$H_B$ & Building height \\
\hline
$K$ & Number of demand points \\
\hline
$R$ & Requested data rate \\
\hline
$\lambda$ & Intensity of UAVs flying along a path \\
\hline
$\lambda_A$ & Intensity of connected UAVs \\
\hline
$h_U$ & UAV altitude \\
\hline
$v_U$ & UAV speed \\
\hline
$P_T$ & Transmit power \\
\hline
$G_T$ & Transmitter's antenna gain \\
\hline
$G_R$ & Receiver's antenna gain \\
\hline
$N_T$ & Number of transmit antenna elements \\
\hline
$N_R$ & Number of receive antenna elements \\
\hline
$L$ & Path loss \\
\hline
$S_T$ & SNR threshold \\
\hline
$N_0$ & Total noise power \\
\hline
$B$ & Available bandwidth \\
\hline
$f_c$ & Carrier frequency \\
\hline
$L_O$ & Maximum connectivity distance \\
\hline
$D_C$ & Distance between connected UAVs \\
\hline
$D_O$ & Distance between disconnected UAVs \\
\hline
$D_C^{\star}$ & Conditional distance between connected UAVs \\
\hline
$\max{D_C}$ & Maximum distance between adjacent connected UAVs \\
\hline
$C$ & Connectivity interval \\
\hline
$D$ & Distance between NR BSs \\
\hline
$R_{\max}$ & Maximum rate supported by the UAV `bridge' \\
\hline
$\zeta$ & Fraction of time when UAV `bridge' exists \\
\hline
$\epsilon$ & Fraction of offloaded data rate \\
\hline
$\nu$ & LoS probability \\
\hline
$T_0$ & Offloading interval length \\
\hline
$T_1$ & Outage interval length \\
\hline
$p_C$ & Connectivity probability between adjacent NR BSs \\
\hline
$p(i,k)$ & Number of turns with random route selection \\
\hline
$p_i$ & Probability of path length $i$ between NR BSs \\
\hline
$W$ & Required data buffer space \\
\hline
\hline
\end{tabular}
\end{table}

\subsection{Deployment and UAV Models}


\begin{figure}[!t]
\centering	
	\subfigure[{Illustration of the offloading process using a UAV `bridge'}]
	{
		\includegraphics[width=0.8\columnwidth]{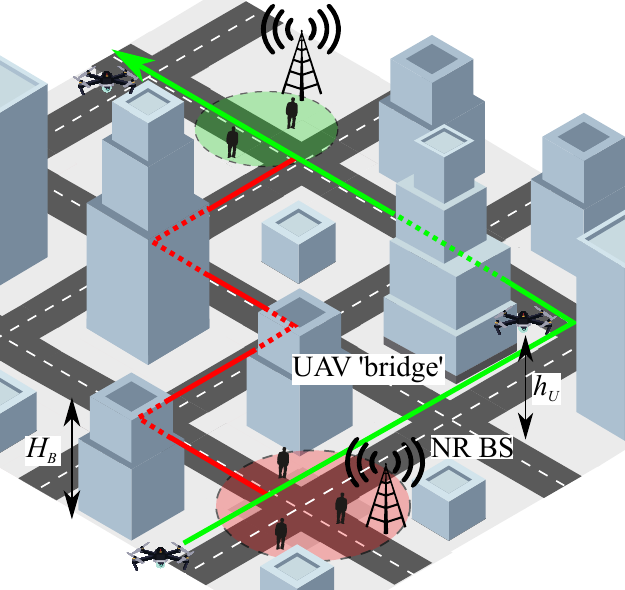}
		\label{fig:drones}
	}\\
	\subfigure[{2D view of the considered deployment}]
	{
		\includegraphics[width=0.8\columnwidth]{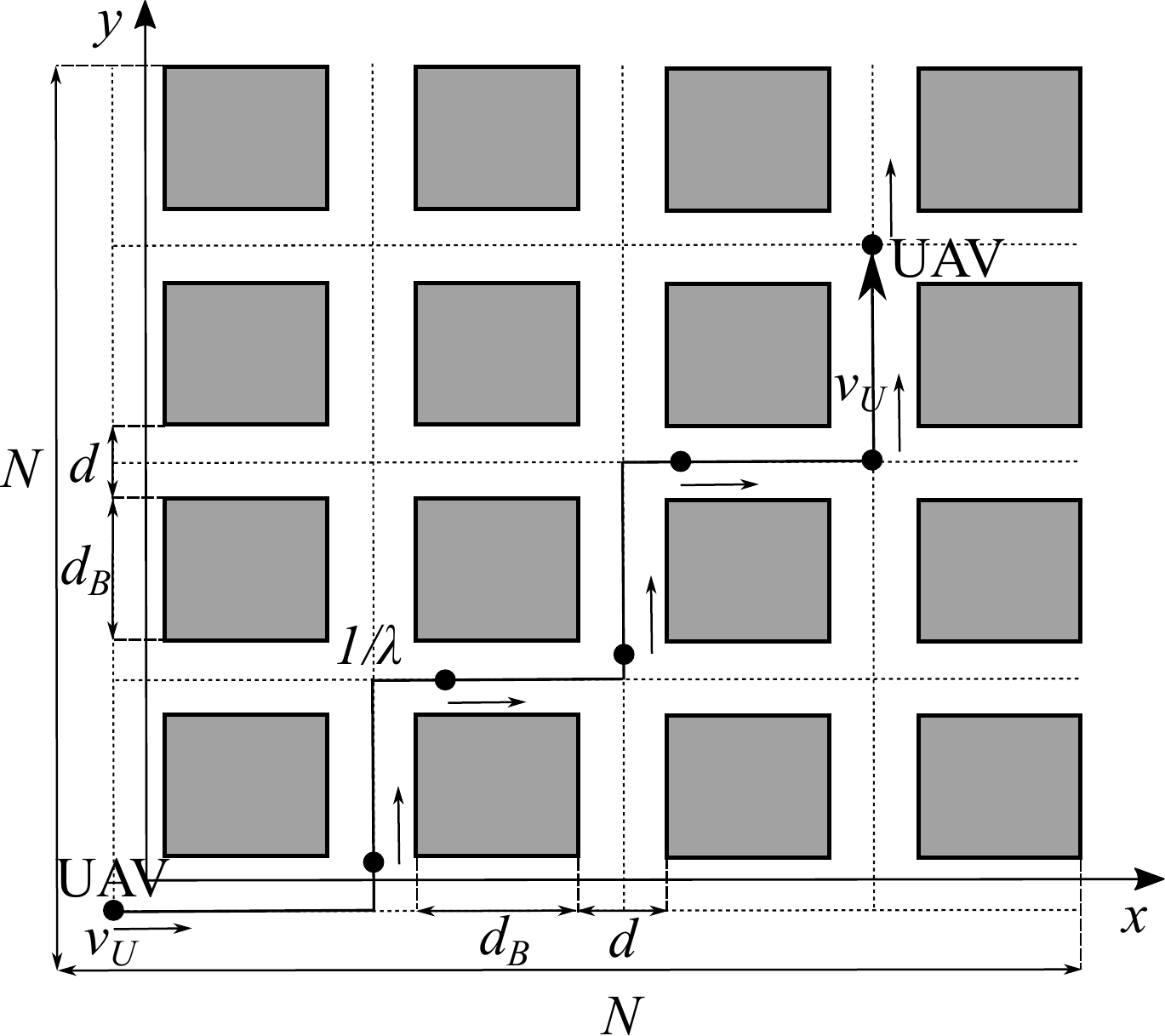}
		\label{fig:2dplot}
	}
\caption{Illustration of the considered mmWave-capable UAV deployment.}
\label{fig:uav_bridge}
\end{figure}

We address a typical scenario for Manhattan-like district as shown in Fig.~\ref{fig:drones} and Fig.~\ref{fig:2dplot}. The area of size $N$ by $N$ blocks consists of streets and buildings with different heights. Lines and squares represent streets and buildings, respectively. The width of the streets is assumed to be constant and equal to $d$. Length of the buildings $d_B$ is also constant. However, building height $H_B$ is a random variable (RV) with the probability density function (pdf) $f_{H_B}(x)$. Note that it is possible to consider various enhancements and modifications to the addressed type of the deployment area. Particularly, one may study non-equal districts having a different number of vertical and horizontal lanes as well as dissimilar width and length of building blocks.


In the considered deployment area, we assume that the user traffic is served by a terrestrial NR infrastructure. However, certain NR BSs may experience overloaded conditions, when the traffic demand exceeds their capacity due to abrupt traffic fluctuations. We assume that at a given time $t$, there are $K$ such \textit{demand source points}. Each demand point is associated with its respective destination, e.g., underloaded NR BS whereto the traffic can be offloaded. Both source and destination NR BSs are assumed to be uniformly distributed over the considered lattice grid. Each demand is characterized by a constant bitrate~$R$.


To offload traffic and overcome potential bottlenecks of limited capacity, we consider mmWave-capable UAVs managed by their UAV operator. In the considered area, the UAVs enter from a corner of the grid and, maintaining their specific route, move to the opposite corner following the homogeneous Poisson processes with intensities $\lambda_i$, $i=1,2,\dots$. The flying altitude $h_U$ and the speed of the UAVs $v_U$ are fixed, and they can only alter their directions in the horizontal plane. The key parameter of interest for the UAV operator is the flight time across the deployment area. Hence, along the path between entry and exit points, only those turns are allowed, which always minimize the flight time.



\subsection{Propagation, Blockage, and Antenna Models}


To model mmWave propagation, we utilize the 3GPP urban micro (UMi) street canyon model. Accordingly, the path loss measured in dB is given by the following:
\begin{align}\label{eqn:pathloss}
\hspace{-3mm}L(f_c, l) = 
32.4 + 21.0 \log_{10} l + 20 \log_{10} f_c,
\end{align}
where $l$ is the distance between the UAVs, $f_c$ is the carrier frequency in GHz. In the considered deployment, the LoS path between the UAVs can be blocked by buildings as the UAVs turn around a building corner. In this case, we assume that the connectivity between the adjacent UAVs is lost.


Various methods may be employed to model antenna arrays at the UAVs, e.g., as in \cite{8290952}. In this paper, we assume that the UAVs are equipped with linear antenna arrays, with the same number of isotropic antenna elements at both endpoints of a connection. Following \cite{constantine2005antenna}, the array factor is defined as a function of the geometry of the array and the phase
\begin{align}\label{eqn:arrayfactor}
F(\theta, \beta) = 
\frac{\sin(M[\pi \cos(\theta)+\beta]/2)}{\sin([\pi \cos(\theta)+\beta]/2)},
\end{align}
where $M$ is the total number of elements in an array, $\beta$ specifies the direction of the array, and $\theta$ is the azimuth angle.
By varying the element separation and the phase between the elements, one can control the properties of the array factor and of the total field of the array. In what follows, we assume $\beta$ = 0 and the array element spacing to be $\lambda_c/2$, where $\lambda_c$ is the wavelength.

Further crucial parameters of the antenna arrays include transmit and receive directivities $\alpha_T$ and $\alpha_R$.  Half-power beamwidth (HPBW) of the array, $\alpha$, is the angular separation over which the magnitude of the radiation pattern decreases by 3 dB. It can be calculated as
\begin{align}\label{eqn:alpha}
\alpha = 2|\theta_m - \theta_{3dB} |,
\end{align}
where $\theta_{3dB}$ is the 3-dB point and $\theta_m$ is the location of the array maximum. The mean antenna gain over HPBW may be computed as
\begin{align}\label{eqn:gain}
G = \frac{1}{\theta^+_{3dB} - \theta^-_{3dB}} \int_{\theta^-_{3dB}}^{\theta^+_{3dB}} \frac{\sin(M\pi \cos(\theta)/2)}{\sin(\pi \cos(\theta)/2} d\theta.
\end{align}


The total received signal power at the UAV is provided as
\begin{align}\label{eqn:SNR}
P_R = P_T + G_T + G_R - L(f_c,l) - N_0(B),
\end{align}
where $P_T$ is the transmit power, $G_T$ and $G_R$ are the antenna gains, $L(f_c, l)$ is the path loss, $N_0(B)$ is the total noise power at the receiver, and $B$ is the bandwidth.


For the SNR threshold, $S_{T}$, the minimal two-dimensional distance, $L_O$, between the adjacent UAVs that results in outage is then computed as
\begin{align}\label{eqn:outage}
L_O = 10^{\frac{P_T + G_T + G_R - N_0 - S_T - 32.4 - 20 \log_{10} f_c}{21}}.
\end{align}

\subsection{Service Strategies with UAV `Bridges'}


For the specified system model, the traffic between the source and the destination demand points is assumed to be offloaded using the UAVs managed by UAV operator according to the rules specified by the SLA. Accounting for value-added nature of the considered service, this has to be done such that the key UAV operator parameter of interest -- the flight time across the deployment area -- is not affected negatively.


In our study, we consider the following offloading schemes:
\begin{itemize}
  \item{\textit{Routed UAV `bridging'.} In this case, the only functionality required from the UAV operator is to navigate a certain fraction of the UAVs along the path between the source and the destination. This is a baseline scheme, which does not call for any advanced SLA between the UAV and the network operators.}
  \item{\textit{Controlled UAV `bridging'.} For this scheme, according to the established SLA between the UAV and the network operators, the latter may not only navigate the UAVs but alter their speed at the entry point of the deployment area, such that the UAV `bridging' performance improves. While this approach may mildly affect the UAV operator's KPIs, the offloading performance may drastically improve.}
\end{itemize}


Utilizing the properties of the considered deployment, there are two crucial improvements to be considered. First, we may introduce additional functionality by allowing the network operator to navigate the UAVs between the source and the destination as long as this does not affect the UAV operator's KPIs in terms of the flight time across the deployment area. Second, even when the number of turns while traversing the deployment area is minimized, potential outage situations may cause significant degradation of the offloading performance. This can be compensated by utilizing additional data buffer space at the UAVs to mitigate shorter service interruptions. In what follows, we consider the introduced schemes with the aforementioned extensions. Therefore, ranging them in the order of the increased complexity of implementation, in terms of the level of control that has to be provided to the network operator, and the additional equipment required at the UAVs (e.g., buffer space), we consider the following strategies: (i) routed with a random route, (ii) routed with route selection, (iii) controlled with a random route, (iv) controlled with route selection, and (v) all aforementioned schemes with buffering.

\subsection{Metrics of Interest}


Note that due to symmetry of the UAVs entry and exit points as well as the flight time requirements of the UAV operator, it is sufficient to consider only one arrival flow of the UAVs, $\lambda_i$. For this arrival flow, the source and the destination nodes are located in the diagonal quadrants. The average intensity of the UAVs serving a single source-destination pair is $\lambda_i/K$, where $K$ is the overall number of NR BSs observing overloaded conditions. For the outlined system model and offloading schemes, we aim to characterize the fraction of traffic demand that can be offloaded onto the UAV `bridge'. Let $U(t)$ be the instantaneous data rate that the UAV `bridge' may support at time $t$. The sought parameter of interest, named the offloading factor, is defined as
\begin{align}
\gamma=\min\{1,\lim_{t\rightarrow\infty}(U(t)/R)\},
\end{align}
where $R$ is the required data rate to be offloaded. In particular, the offloading factor indicates the portion of traffic demand that can be offloaded, if the UAV `bridge' throughput is lower than the amount of traffic at the NR BS. Therefore, the offloading factor is upper bounded by one, i.e., when the offloading factor is strictly one, this implies that all the traffic can be offloaded. Hence, taking it to the limit and upper bounding by one delivers the fraction of traffic that is offloaded.

\section{Analysis of Offloading Strategies}\label{sect:framework}

In this section, we first describe the developed methodology. Then, we proceed by characterizing the UAV `bridge' connectivity properties and finally analyze the offloading strategies.

\subsection{Approach at a Glance}

We start with the routed UAV `bridging' scheme. First, we characterize the UAV `bridge' connectivity properties by obtaining the fraction of time when there is a multi-hop relay link between the source and the destination NR BSs, $\zeta$. To achieve that, we estimate the distributions of connectivity intervals formed by consecutive UAVs and distance between the source and the destination NR BSs. Then, we define the offloading interval as the difference between these two. We further estimate the maximum supported data rate at the UAV `bridge', $R_{\max}$, as a function of the maximum distance between the UAV forming the `bridge'. Using the requested data rate $R$, this will provide the fraction of demand that can be offloaded onto the `bridge' as $\epsilon=\min\{1,R_{\max}/R\}$. Finally, we account for the effect of mmWave LoS blockage by estimating the probability, $1-\nu$, that a route between the source and the destination does not exist due to blockage at street intersections. Finally, the offloading factor is readily provided as $\gamma=\zeta\epsilon\nu$.

We also address the controlled UAV `bridging' scheme. The main difference is that $\zeta$ becomes the step function, which accepts zero for all the UAV intensities $\lambda$ lower than a certain value. The rest of the derivations remain the same. Finally, by using the derived parameters $\zeta$, $\nu$, and $\epsilon$, we obtain the offloading factors for both routed and controlled UAV `bridging' schemes with introduced enhancements. The crucial step at this stage is to determine the amount of buffer space required to alleviate the effect of LoS blockage at the intersections and determine the number of intersections along the path between the source and the destination NR BSs.

\subsection{UAV `Bridge' Process Characterization}


We begin by characterizing the connectivity process between the source and the destination NR BSs that can be represented by the temporal Poisson process with a certain intensity $\lambda$ of UAVs traveling along the same path. Observe that the connectivity process depends on two main factors: the distance between the UAVs and the LoS blockage by buildings at the intersections. Below, we first characterize the offloading interval induced by the distances between the adjacent UAVs and then superimpose it with the LoS link blockage.

\subsubsection{Fraction of Offloading Time, $\zeta$}


To determine the offloading interval between the source and the destination NR BS induced by the distances between the adjacent UAVs, we first determine the length of the connectivity interval on a line; then, we obtain the distance between the source and the destination NR BSs. Finally, we determine the difference between them. Recall that the distance between two points of the one-dimensional Poisson process having intensity $\lambda$ follows an exponential distribution with parameter $\lambda$. Given the maximum connectivity distance of UAV, $L_O$, as defined in (\ref{eqn:outage}), the pdf of the distance between two connected UAVs, $D_C$, obeys
\begin{align}
f_{D_C}(x)=\frac{\lambda{}e^{-\lambda{}x}}{\int_{0}^{L_O}\lambda{}e^{-\lambda{}y}dy}=\frac{\lambda e^{-\lambda x}}{1-e^{-\lambda L_O}},\,0<x<L_O.
\end{align}

\begin{figure}[!t]
\centering	
	\subfigure[{Offloading interval}]
	{
		\includegraphics[width=0.8\columnwidth]{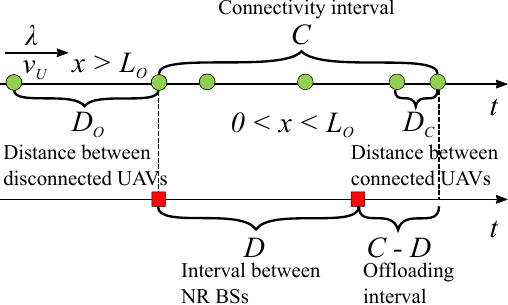}
		\label{fig:connectivity_1}
	}\\
	\subfigure[{Interval when offloading is infeasible}]
	{
		\includegraphics[width=0.8\columnwidth]{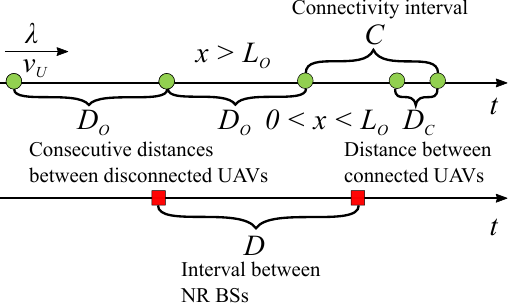}
		\label{fig:connectivity_2}
	}
\caption{Illustration of intervals for fraction of offloading time.}
\label{fig:uav_connectivity}
\end{figure}

Observe that the connectivity interval can be formed by several UAVs, whose inter-UAV distance is less than the outage distance $L_O$ (Fig. \ref{fig:connectivity_1}). Since successive intervals in the Poisson process are independent RVs, the number of successive inter-UAV intervals forming a connectivity interval follows a geometric distribution with the parameter
\begin{align}\label{eqn:pC}
p_C=\int_{0}^{L_O}\lambda{}e^{-\lambda{}y}dy=1-e^{-\lambda{}L_O}.
\end{align}

Let $C$ denote the length of the connectivity interval. This RV can be written as a sum of inter-UAV distances conditioned on the fact that it is smaller than $L_O$ and weighted with geometric coefficients $p_C^{i}(1-p_C)$, $i=1,2,\dots$. Therefore, the pdf of the connectivity interval can be written as
\begin{align}\label{eqn:connInt0}
f_{C}(x)=\sum_{i=1}^{\infty}p_C^{i}(1-p_C)\left[f_{D_C}\underset{i\text{ times}}{\star{}}f_{D_C}\right](x),
\end{align}
where $\star$ denotes a convolution operation, i.e.,
\begin{align}\label{eqn:convDC}
[f_{D_C}\star{}f_{D_C}](x)=\hspace{-2mm}\int\limits_{0}^{2L_O}\hspace{-1mm}f_{D_C}(x-\tau)f_{D_C}(\tau)d\tau.
\end{align}


Note that (\ref{eqn:connInt0}) cannot be evaluated in the closed-form. To provide a suitable approximation, one may replace the bounded exponential distribution $f_{D_C}(x)=\lambda{}e^{-\lambda{}x}/(1 - e^{-\lambda{}L_O})$ by an exponential distribution with the appropriate parameter,
\begin{align}\label{eqn:lambdaA}
\lambda_{A}=\frac{1}{\int_{0}^{L_O}\frac{\lambda}{e^{-\lambda{}L_O}}e^{-\lambda{}x}xdx}=\left[\frac{1}{\lambda }-\frac{L_O}{e^{\lambda  L_O}-1}\right]^{-1}.
\end{align}

Further, it may be observed that the convolution of $i$ exponentially distributed RVs with the same parameter is Erlang distribution of order $i$, $E(i,\lambda_A)$, with the pdf 
\begin{align}
f_{E(i,\lambda_A)}(x)=\frac{\lambda_A^{i}}{(i-1)!}x^{i-1}e^{-\lambda_A{}x},\,x>0,\lambda_A>0,
\end{align}
while the result in (\ref{eqn:connInt0}) can be approximated by
\begin{align}\label{eqn:connIntFinal}
\hspace{-2mm}f_{C}(x)&=\sum_{i=1}^{\infty}(1-e^{-\lambda{}L_O})^{i-1}e^{-\lambda L_O}\frac{\lambda_A^{i}x^{i-1}e^{-\lambda_A{}x}}{(i-1)!}=\nonumber\\
\hspace{-2mm}&=e^{-\lambda_A{}x-\lambda L_O}\sum_{i=1}^{\infty}\frac{\lambda_A^{i}x^{i-1}}{(i-1)!}(1-e^{-\lambda L_O})^{i-1}=\nonumber\\
\hspace{-2mm}&=\frac{\lambda  [e^{\lambda  L_O}-1]e^{-\frac{\lambda  [\lambda  L_O^2-L_O e^{\lambda  L_O}+x[e^{-\lambda L_O}-1]+L_O]}{\lambda L_O-e^{\lambda  L_O}+1}}}{e^{\lambda  L_O}-\lambda L_O-1}.
\end{align}


Observe that there is a connected multi-hop relay between the source and the destination NR BSs, only if the Manhattan distance between them is smaller than the length of the connectivity interval, $C$. Let RV $D$ denote the Manhattan distance between a source-destination demand pair, which is readily given by the sum of coordinate differences of the source and the destination, $(X_1,Y_1)$ and $(X_2,Y_2)$, i.e.,
\begin{align}
|X_1-X_2|+|Y_1-Y_2|,
\end{align}
where $X_i$ and $Y_i$ are uniformly distributed in $(0,N)$.

The difference between two identical uniform distributions is given by a triangular distribution with parameters $(-N,N)$:
\begin{align}\label{eqn:d_sub}
p_{X_1-X_2}(i) = p_{Y_1-Y_2}(i) = \frac{N+1-|i|}{(N+1)^2},i\leq{}|N|.
\end{align}

Since this distribution is symmetric, the distributions of $|X_1-X_2|$ and $|Y_1-Y_2|$ are both defined over $(0,N)$, and they are twice the sum of the positive branch of (\ref{eqn:d_sub}), i.e.,
\begin{align}\label{eqn:d_abs}
p_{|X_1-X_2|}(i) =
\begin{cases}
\displaystyle{\frac{1}{N+1}},&i=0,\\
\displaystyle{\frac{2(N+1-i)}{(N+1)^2}},&0<i\leq{}N.
\end{cases}
\end{align}

The probability mass function (pmf) of $|X_1-X_2|+|Y_1-Y_2|$ therefore reads as in (\ref{eq:distancepmf}).

\begin{figure*}[!!t]
\vspace{-0mm}
	\begin{align}\label{eq:distancepmf}
p_i=
\begin{cases}
\displaystyle{\frac{1}{(N+1)^2},}&i=0,\\
\displaystyle{\frac{4(N+1-i)}{(N+1)^2}+\frac{(-1)^i+1}{2}\frac{4(N+1-i/2)^2}{(N+1)^4}+\sum^{\lfloor{}i/2-1\rfloor{}}_{k=1}\frac{4(N+1-k)(N+1-i-k)}{(N+1)^4}},&i=1,2,\dots,2N.\\
\end{cases}
	\end{align}
\hrulefill
\normalsize
\vspace{-3mm}
\end{figure*}




After characterizing the connectivity interval and the distance between the source and the destination NR BSs, we may proceed with determining the mean length of the offloading interval when multi-hop communication is feasible. The pdf of the sought RV, $T_0$, is provided by
\begin{align}
f_{T_0}(x)=
\begin{cases}
\displaystyle{f_{C-D}(x),}&x>0,\\
\displaystyle{\int_{-\infty}^{0}f_{C-D}(x)dx},&x=0.\\
\end{cases}
\end{align}

One can determine the pdf of $T_0$ by a convolution of $C$ and $(-D)$. Owing to the discrete nature of $D$, we have the following for cumulative distribution function (CDF) of $C-D$
\begin{align}
F_{C-D}(x)=\sum_{j=0}^{i}F_{C}(x+j)p_{j},
\end{align}
where the CDF of $C$ is obtained from (\ref{eqn:connIntFinal}) by integration.


Observe that the length of the interval where no multi-hop communication is available between the source and the destination NR BS, $E[T_1]$, is obtained similarly. The difference is that here, we need to consider the consecutive distances between the UAVs, which are greater than the outage distance $L_O$ (Fig. \ref{fig:connectivity_2}). Consequently, the fraction of time when the relay link is available is given by
\begin{align}\label{eqn:linkExists}
\zeta=\frac{E[T_0]}{E[C]+E[T_1]},
\end{align}
which is independent of the UAVs velocity, $v$.

The primary difference between the routed and controlled UAV `bridging' is in that with the latter scheme the UAV operator may delay the UAVs at the entry point, such that the distance between them is constant and equals to $1/\lambda$. Hence, the UAV `bridge' of infinite length exists, i.e., $\zeta=1$, when $1/\lambda<L_O$. Otherwise, $\zeta=0$ and no `bridging' is feasible.

\subsubsection{Fraction of Offloaded Rate, $\epsilon$}


Once we established the fraction of time when a multi-hop relay link is available, we determine the fraction of data traffic offloaded onto this relay link. The latter at any instant of time when the link is available is given by $\epsilon=\min\{1,R_{\max}/R\}$, where $R_{\max}$ is the maximum data rate provided by the UAV `bridge' and $R$ is the constant requested rate. As the latter parameter is a part of the system specification, the only unknown is $R_{\max}$.


Let us now assess the data rate provided by the UAV `bridge', $R_{\max}$. Observe that for the multi-hop UAV `bridge' consisting of $i$ UAVs, the data rate is upper bounded by the minimum rates of two adjacent UAVs. The latter is obtained as the link having maximum communications distance out of all $i-1$ links. Hence, one needs to determine the maximum of $i-1$ exponential RVs and then weigh them with the probabilities of having exactly $i-1$ UAVs in the `bridge', $p_{C}^{i-1}(1-p_{C})$. The CDF of the maximum of $i$ independent exponentially distributed RVs is given by \cite{ross2014introduction}
\begin{align}\label{eqn:max}
f_{\max{D_C}}(x;i-1)&=(i-1)f_{D_C}(x)[F_{D_C}(x)]^{i-2}=\nonumber\\
&=(i-1)\lambda_Ae^{-\lambda_{A}x}(1-e^{-\lambda_Ax})^{i-2},
\end{align}
where $\lambda_A$ is provided in (\ref{eqn:lambdaA}).

Further, the pdf of the maximum link distance is
\begin{align}\label{eqn:maxD}
&f_{\max{D_C}}(x)=\sum_{i=1}^{\infty}p_{C}^{i-1}(1-p_{C})f_{\max{D_C}}(x;i-1)=\nonumber\\
&=\sum_{i=1}^{\infty}\frac{(1-e^{-\lambda{}L_O})^{i-1}(i-1)\lambda_A(1-e^{-\lambda_Ax})^{i-2}}{e^{\lambda{}L_O+\lambda_Ax}}=\nonumber\\
&=\frac{\lambda  (e^{\lambda  L_O}-1) e^{\lambda  \left(\frac{\lambda  L_O x}{e^{\lambda  L_O}-1}-L_O\lambda+L_O+x\right)}}{\left(e^{\lambda  L_O}-1-L_O\lambda\right) \left(e^{\lambda  L_O}+e^{\frac{x}{\frac{1}{\lambda }-\frac{L_O}{e^{\lambda  L_O}-1}}}-1\right)^2}.
\end{align}


Once the maximum link distance between the adjacent UAVs is determined, we employ the non-linear RV transformation technique to determine the pdf of the maximum data rate supported by the UAV `bridge', $f_{R_{\max}}(x)$. Particularly, recall that the pdf of the RV $Y$, $w(y)$, which is expressed as function $y=\phi(x)$ of another RV $X$ with the pdf $f(x)$, is given by
\begin{align}\label{eqn:rvTrans}
w(y)=\sum_{\forall{i}}f(\psi_i(y))|\psi_i{'}(y)|,
\end{align}
where $x=\psi_i(y)=\phi^{-1}(x)$ are the inverse functions.

The inverse of the data rate function and its derivative are
\begin{align}\label{eqn:deriv}
&\psi(y)=5.53\times 10^6\left[\frac{B f_c^2 10^{-\frac{P_T}{10}} (2^{\frac{x}{B}}-1)}{N_R N_T}\right]^{-\frac{10}{21}},\nonumber\\
&|\psi{'}(y)|=\frac{1.82\times 10^6 f_c^2 e^{-0.23 P_T} 2^{\frac{x}{B}}}{N_R N_T \left(\frac{B f_c^2 e^{-0.23 P_T} \left(2^{\frac{x}{B}}-1\right)}{N_R N_T}\right)^{\frac{31}{21}}}.
\end{align}

Substituting (\ref{eqn:deriv}) into (\ref{eqn:maxD}), we arrive at the pdf of the maximum data rate supported by UAV `bridge' in the closed-form, $f_{R_{\max}}(x)$. Now, the pdf of $R_{\max}/R$ is obtained by a straightforward scaling of $R_{\max}$, i.e., $f_{R_{\max}}(xR)$. Further, the pdf of $\min\{1,R_{\max}/R\}$ is given by 
\begin{align}
f_{\min\left\{1,\frac{R_{\max}}{R}\right\}}(x)=
\displaystyle{\begin{cases}
f_{R_{\max}}(xR),&0<x<1,\\
\int_{1}^{\infty}f_{R_{\max}}(yR)dy,&x=1.\\
\end{cases}}
\end{align}

Finally, the fraction of the offloaded data rate at any given instant of time when there exists a UAV `bridge' between the source and the destination NR BSs is provided by
\begin{align}\label{eqn:fractionTraffic}
\epsilon=\int_{0}^{1}f_{\min\{1,R_{\max}/R\}}(x)xdx.
\end{align}

For the controlled `bridging' scheme, derivation of $\epsilon$ is similar to the routed scheme, except for a lack of randomness in the inter-UAV distance.


\begin{figure}[t!]
\vspace{-0mm}
\centering
\includegraphics[width=0.9\columnwidth]{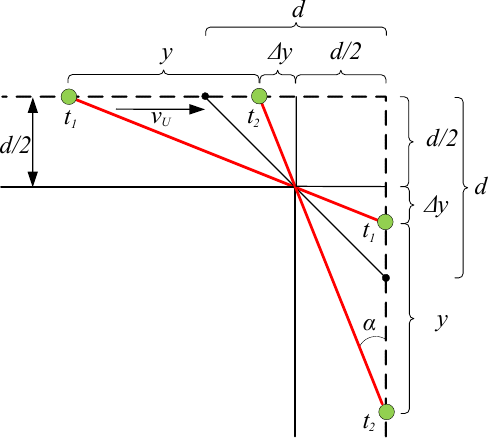}
\caption{Blockage of mmWave LoS path at an intersection.}
\label{fig:corner}
\vspace{-3mm}
\end{figure} 

\subsubsection{LoS Probability, $\nu$}


In addition to the intensity of the UAVs along the path between a source and a destination, buildings may also affect the connectivity process. In particular, as illustrated in Fig. \ref{fig:corner}, at intersections they can occlude the LoS path between the adjacent UAVs in a `bridge'. In this case, the `bridge' between the source and the destination NR BSs cannot be established even when the distance between the adjacent UAVs allows it. The corresponding probability, $\nu$, depends on the number of intersections along the path between the source and the destination NR BSs as well as the height of the building blocks.


Let $q_i$ be the probability that there are $i$ turns along the path between the source and the destination NR BSs. Since only those steps that minimize the distance to the exit points are preferred by the UAV operator, the numbers of even and odd intersections crossed by the UAV when passing from $(0,0)$ to $(N,N)$ are given by \cite{kuo2016enumeration}
\begin{align}
&p(N,2k)=2{N-1 \choose k-1}{N-1 \choose k},\nonumber\\
&p(N,2k-1)=2{N-1 \choose k-1}^2.
\end{align}

Accounting for a random distance between the source and the destination NR BSs as considered in (\ref{eq:distancepmf}), the probability of crossing exactly $i$ intersections can be established by $q_{i}=p(i,k)p_{i}$. Using the intersection geometry illustrated in Fig. \ref{fig:corner}, we further observe that no LoS blockage occurs when the height of the building, $H_B$, is greater than the UAV altitude, $h_U$, and the distance between the UAVs is less than $d$, where $d$ is the street width. Hence, the relay process is interrupted when the distance between the connected UAVs is within $D_C^{\star}\in(d,L_O)$. The pdf of $D^{\star}_C$ is given by
\begin{align}\label{eqn:Dstar}
f_{D^{\star}_C}(x)&=\frac{f_{D}(x)}{1-[F_{D}(L_O)-F_D(d)]}=\nonumber\\
&=\frac{1}{e^{-d \lambda }-e^{-\lambda  L_O}}\lambda  e^{-\lambda x},\,d<x<L_O.
\end{align}

Therefore, the LoS probability constitutes
\begin{align}\label{eqn:LoSblockfinal}
\nu=\sum_{i=1}^{4N}p(i,k)p_{i}\left(F_{H_B}(h_U)\left[1-\int\limits_{d}^{L_O}f_{D_C^\star}(x)dx\right]\right)^i,
\end{align}
where $h_U$ is the UAV altitude. 

Substituting $F_{H_B}(x)$, $f_{D^{\star}_C}(x)$, as well as $p_i$ and $p(i,k)$ derived earlier into (\ref{eqn:LoSblockfinal}), we obtain $\nu$. For the fully controlled `bridging' scheme, the rest of the analysis is analogous to the routed scheme, except for the fact that the inter-UAV distance is not random.

\subsection{Assessing Considered Enhancements}

In this subsection, we address two enhancements to the proposed `bridging' schemes that may drastically improve the offloading gains. These are (i) minimizing the number of intersections along the path between the source and the destination NR BSs and (ii) the use of buffering at the UAVs to mitigate the effects of LoS blockage.

\subsubsection{Route Selection}

One of the straightforward ways to decrease the disruptions caused by the LoS blockage is to enable route selection within the considered deployment area. Observe that between two arbitrarily distributed points on a lattice grid there always exists a path having only one turn. In this case, the LoS blockage probability reduces to
\begin{align}
\nu=F_{H_B}(h_U)\left[1-\int_{d}^{L_O}f_{D_C^\star}(x)dx\right],
\end{align}
which can be evaluated in the closed-form for a given $f_{H_B}(x)$.

\subsubsection{Buffering at UAVs}

Since UAVs are expected to fly at relatively high speeds reaching $60-80$ km/h ($16-22$ m/s), the `bridge' connectivity interruptions caused by LoS blockage can be alleviated by using additional buffer space at the UAVs. Below, we estimate the required amount of such space needed to smoothen the harmful effects.

Let $W$ denote the required buffer space in order to alleviate the effects of blockage. We thus have
\begin{align}\label{eqn:buffer}
W=R_{\max}Y/v_U,
\end{align}
where $Y$ is the RV denoting the time that the UAV spends in the LoS blocked conditions, $R_{\max}$ is the maximum data rate provided by the UAV `bridge', and $v_U$ is the UAV velocity.

To determine $Y$, we first observe (Fig. \ref{fig:corner}) that
\begin{align}
\tan\alpha=\frac{\Delta{y}+d/2}{y+\Delta{y}+d/2}=\frac{\Delta{y}}{d/2},
\end{align}
which results in the following equation with respect to $\Delta{y}$
\begin{align}
(\Delta{y})^2+(d-D_C)\Delta{y}+d^2/4=0.
\end{align}
The respective solution of interest is
\begin{align}
\Delta{y}=(\sqrt{D_C^2-2D_Cd}+D_C-d)/2.
\end{align}

Once $\Delta{y}$ is found, the LoS blockage distance is derived as
\begin{align}\label{eqn:igrek}
Y=(\sqrt{D^{\star2}_C-2D^{\star}_Cd}+D^{\star2}_C-d)/2,
\end{align}
where $D^{\star}_C$ is the conditional distance between two adjacent UAVs, whose pdf is provided in (\ref{eqn:Dstar}).

Substituting (\ref{eqn:igrek}) into (\ref{eqn:buffer}), the RV of interest is
\begin{align}\label{eqn:buffer2}
W=\frac{R_{\max}(\sqrt{D^{\star2}_C-2D^{\star}_Cd}+D^{\star2}_C-d)}{2v_U}.
\end{align}

The pdf of $W$, $f_W(x)$, which is produced according to (\ref{eqn:buffer2}), can be established by using non-linear RV transformation technique \cite{ross2014introduction}. To assess the mean buffer size requirements at the UAVs, it is sufficient to apply Taylor series expansion, similarly to \cite{kovalchukov2018evaluating}.

\section{Numerical Results}\label{sect:numerical}

In this section, we numerically elaborate on the performance of the proposed offloading strategies. First, we assess the accuracy of the developed mathematical models for the considered offloading schemes. Then, we investigate the operation of our offloading strategies as functions of the system parameters, which include the deployment area, the UAV characteristics, and the radio settings. Here, we also assess the volume of buffer space required at the UAVs to efficiently mitigate the effect of blockage. Finally, we compare the proposed approach to that of the standardized baseline IAB-based scheme introduced by 3GPP. The default system modeling parameters are provided in Table~\ref{tab:parameters}.

\begin{table}[t!]
\caption{Default system parameters for numerical assessment.}
\label{tab:parameters}
\footnotesize
\begin{tabular}{p{0.55\columnwidth}p{0.3\columnwidth}}
\hline
\textbf{Parameter}&\textbf{Value} \\
\hline\hline
Carrier frequency, $f_c$ & 73\,GHz \\
\hline
Bandwidth, $B$ & 100\,MHz \\
\hline
Default number of streets, $N$ & 10 \\
\hline
Default deployment type & Dense urban \\
\hline
Street width, $d$ & 10/13/20/20\,m \\
\hline
Building block length, $d_B$ & 37/45/60/60\,m \\
\hline
Building height, $H_B$ & 10/19/25/63\,m \\
\hline
Transmit power, $P_T$ & 23\,dBm \\
\hline
SNR threshold, $S_T$ & 10\,dB \\
\hline
Number of transmit antenna elements, $N_T$ & 4 \\
\hline
Number of receive antenna elements, $N_R$ & 4 \\
\hline
UAV altitude, $h_U$ & 30\,m \\
\hline
UAV speed, $v_U$ & 10\,m/s \\
\hline
Requested data rate, $R$ & 3\,Gbps \\
\hline
Default intensity of UAVs, $\lambda$ & 0.04\,1/m \\
\hline
\hline
\end{tabular}
\vspace{-0mm}
\end{table}

\begin{figure}[!t]
    \centering
    \includegraphics[width=0.75\columnwidth]{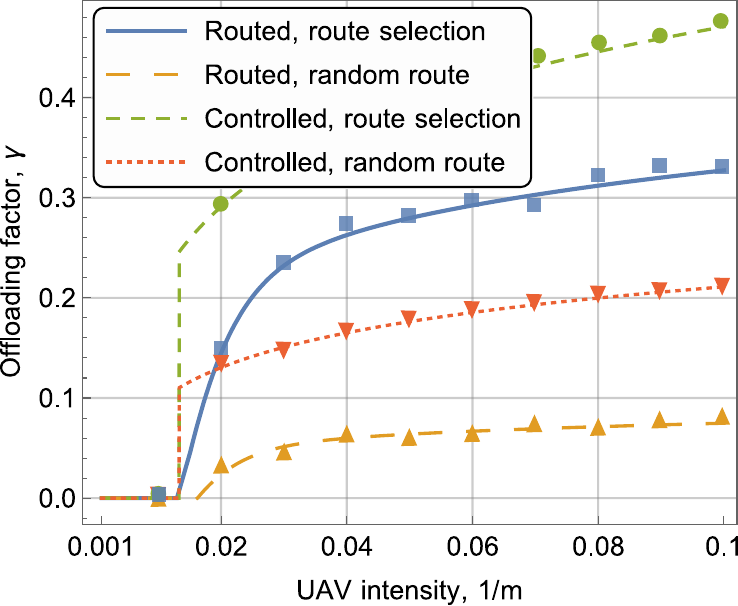}
    \caption{Comparison of our model and SLS results.}
    \label{fig:sls}
    \vspace{-0mm}
\end{figure}

\subsection{Model Validation}

We first validate our model by using a custom-made system-level simulator (SLS) named WINTERsim, which has been extensively utilized in the past for 5G/5G+ performance evaluation \cite{7084578,mmWaveJsac,petrov2018achieving}. In this simulator, based on the latest 3GPP specifications, the PHY and MAC layers are implemented in detail, while the upper layers are simplified to abstract the traffic models supported with analytical approximations. Our SLS tool supports 3D geographical models, which take into account interference patterns and antenna configurations.

For our simulation campaign, we utilized the system input variables from Table \ref{tab:parameters} and modeled the offloading factor for two basic schemes, routed and controlled, each having two operating modes -- route selection and random route choice. In the considered scenario, UAVs are generated in the upper-left corner according to a Poisson process with the intensity of~$\lambda$. They move along the streets towards the source NR BS. Then, depending on the selected strategy, we route them with either a single turning point or a random route is generated towards the destination NR BS. To account for any uncertainties in the NR BS locations and to collect statistical data, the method of replications is used \cite{perros2009computer}, i.e., in each simulation run the positions of the source and the destination NR BSs are generated randomly. The number of replications is set to 1000. Further, in each replication, 1000 observations are collected.

A comparison of the simulation and analysis results is demonstrated in Fig. \ref{fig:sls} as a function of the UAV intensity. As one may observe, there is an excellent match between the modeled data and the simulations. At the same time, smaller inconsistencies in the tightness of convergence may be noticeable. The reason is that the routed schemes add another level of stochasticity to the random distances between the adjacent UAVs for a given intensity of the UAVs entering the area. However, even for this scheme, the absolute deviations are rather small, which implies that the developed models allow to accurately capture the system performance. For this reason, in the rest of this section, we rely on the developed mathematical model to characterize the offloading factor over a wide range of the input system parameters.

\subsection{Comparison of Offloading Schemes}


We start our evaluation campaign by comparing the performance of the analyzed offloading schemes as a function of the system parameters and deployment choices. Following the International Telecommunication Union (ITU) radiocommunication unit's specification \cite{standard_itu}, we differentiate between suburban, urban, dense urban, and highrise urban deployments. Each deployment type is characterized by a certain mean width of streets, $d$, mean length of building blocks, $d_B$, as well as mean height of the buildings, $E[H_B]$, as a function of population density in various deployment types provided in \cite{standard_itu}. Accordingly, the building height distribution is assumed to follow a Rayleigh distribution with the parameter $\sigma=E[H_B]/\sqrt{\pi/2}$.

\begin{figure}[!t]
\centering	
	\subfigure[{Effect of UAV intensity}]
	{
		\includegraphics[width=0.75\columnwidth]{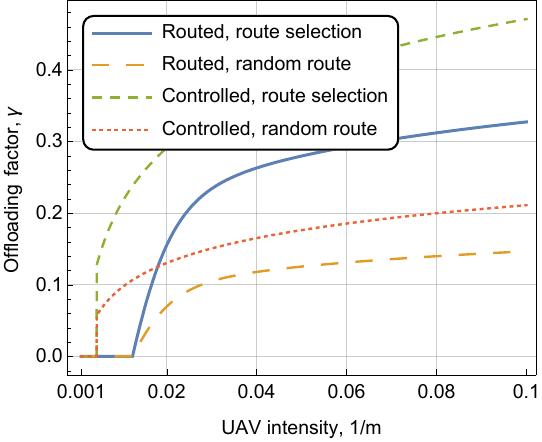}
		\label{fig:1Lambda}
	}\\
	\subfigure[{Effect of deployment area}]
	{
		\includegraphics[width=0.75\columnwidth]{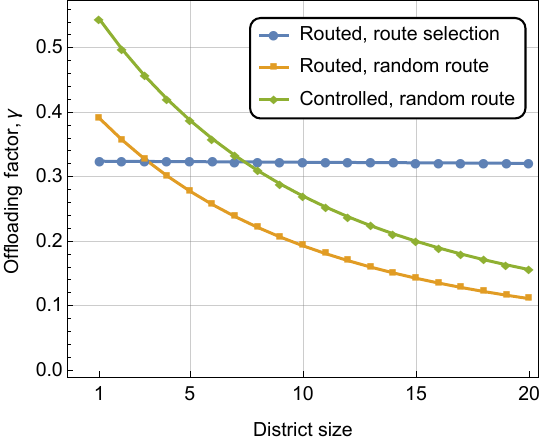}
		\label{fig:1N}
	}
\caption{Effects of UAV intensity and deployment size.}
\label{fig:1LambdaN}
\end{figure}


First, Fig. \ref{fig:1LambdaN} illustrates the impact of the UAVs intensity on the offloading factor for the routed and controlled offloading schemes, dense urban deployment, $4\times{}1$ transmit and receive antenna arrays, $N=10$ vertical and horizontal streets, $R=3$ Gbps, $E[H_B]=25$ m, $h_U=30$ m. Analyzing the data in Fig. \ref{fig:1Lambda}, one may observe that the routed scheme allows for establishing a UAV `bridge' starting from the UAV intensity of approximately $0.01$, which corresponds to the inter-UAV distance of around $100$ m. Starting from that point, the offloading factor grows rapidly. Note that the use of the route selection functionality, which results in choosing routes between the source and the destination NR BSs with only one turn, leads to a dramatic improvement in terms of the offloading gains, with the difference reaching $0.1$ for moderate to high UAV intensities. Observe that this scheme may even outperform its controlled counterpart (without route selection functionality) starting from the UAV intensity of approximately $0.02$.


Analyzing the data in Fig. \ref{fig:1Lambda} further, one may notice that the controlled schemes may operate under a wider range of UAV intensities. Particularly, the offloading starts already with the UAV intensity of approximately $0.008$, which corresponds to $125$ m. Furthermore, the gains of route selection are also much higher as compared to the routed scheme, and reach $0.2$. This is explained by the stochastic nature of the routed scheme, where any UAV `bridge' interruption as a result of the LoS blockage is determined by the maximum inter-UAV distance. The mean of this distance is higher than the mean distance between the adjacent UAVs. Hence, we may conclude that forming deterministic UAV `bridges' by delaying UAVs at their entry point leads to significant offloading gains.

\begin{figure}[!t]
\centering	
	\subfigure[{Routed with route selection}]
	{
		\includegraphics[width=0.75\columnwidth]{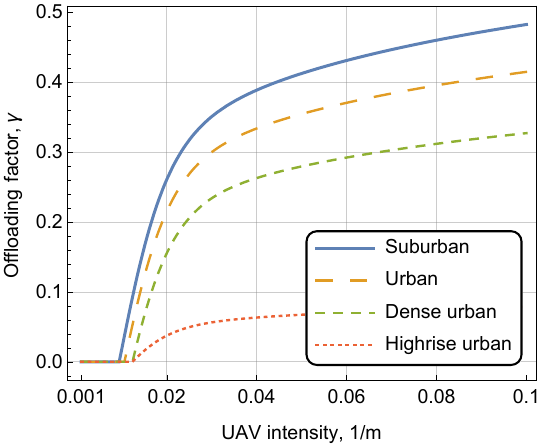}
		\label{fig:1_districtRouted}
	}\\
	\subfigure[{Controlled with route selection}]
	{
		\includegraphics[width=0.75\columnwidth]{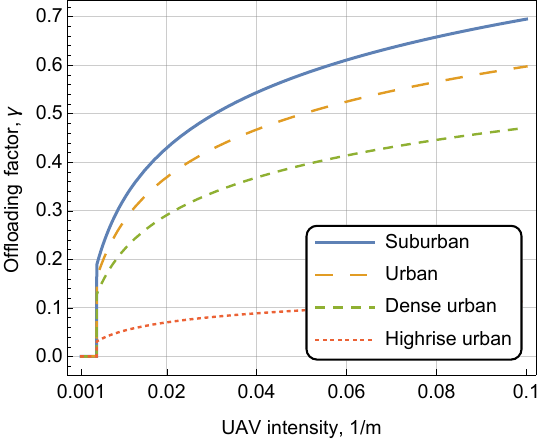}
		\label{fig:1_districtControlled}
	}
\caption{Effects of deployment type.}
\label{fig:1_disctrict}
\end{figure}


Consider now the effects of deployment size, $N$, as illustrated in Fig. \ref{fig:1N}. As one may observe, an increase in $N$ leads to lower offloading factor for the schemes without route selection. Indeed, higher values of $N$ lead to a higher distance between the source and the destination NR BSs, thus increasing the number of turns along the path. This negatively affects the fraction of time when there exists a UAV `bridge', which yields a small offloading factor. Note that this effect is not observed when route selection is enabled, thus ensuring that at most one turn along the path between the source and the destination NR BSs is performed. Therefore, the impact of longer path associated with increasing $N$ is almost negligible as compared to the number of turns.



We now consider the influence of the deployment area on the offloading factor by implicitly specifying the width of buildings and streets as well as the average building height. Fig. \ref{fig:1_disctrict} shows the offloading factor for the routed and controlled schemes with route selection as a function of the UAV intensity, $\lambda$, $N=10$, $h_U=30$ m, and $4\times{}1$ transmit and receive antenna arrays. As one may learn, higher offloading factor is achieved for suburban deployments, while highrise urban deployment is characterized by the poorest performance. This is explained by the mean building height, which is the tallest for highrise urban layouts. Furthermore, this fact is also amplified by the width of streets and building blocks, which are the smallest for suburban areas and the biggest for highrise urban deployments.

\begin{figure}[!t]
    \centering
    \includegraphics[width=0.75\columnwidth]{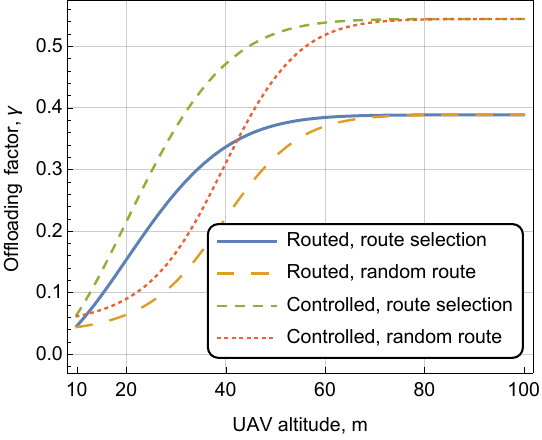}
    \caption{Impact of UAV altitude.}
    \label{fig:uav_alt}
    \vspace{-0mm}
\end{figure}

The impact of the UAV altitude for dense urban deployment type on the offloading factor for the routed and controlled schemes is illustrated in Fig. \ref{fig:uav_alt} for $\lambda=0.04$ 1/m, $N=10$, $E[H_B]=25$ m, and $4\times{}1$ transmit and receive antenna arrays. Here, increasing the UAV altitude improves the offloading performance for both routed and controlled schemes. Importantly, as $h_U$ grows, both schemes without route selection approach the performance of those with route selection. The reason is that even with a higher number of turns along the path between the source and the destination NR BSs, the probability of at least one LoS blockage decreases drastically. Hence, to alleviate the negative effects of blockage, one may either enforce route selection or increase the UAV altitude. However, to fully mitigate performance degradation, additional efforts may be required. 

\subsection{Effects of Buffering}


As discussed in the previous subsection, blockage by buildings may have a profound effect on the performance of the offloading schemes. Even though a careful choice of routes between the source and the destination NR BSs allows to improve the system performance, there is still a non-negligible chance that the route is disrupted by an occluding building. This disadvantage can be alleviated by utilizing additional buffer space at the UAVs that will be used to temporarily store data while the LoS conditions are interrupted between the adjacent UAVs. This should improve the offloading gains of the considered schemes at the expense of utilizing additional storage space.

\begin{figure}[!t]
\centering	
	\subfigure[{Routed scheme}]
	{
		\includegraphics[width=0.75\columnwidth]{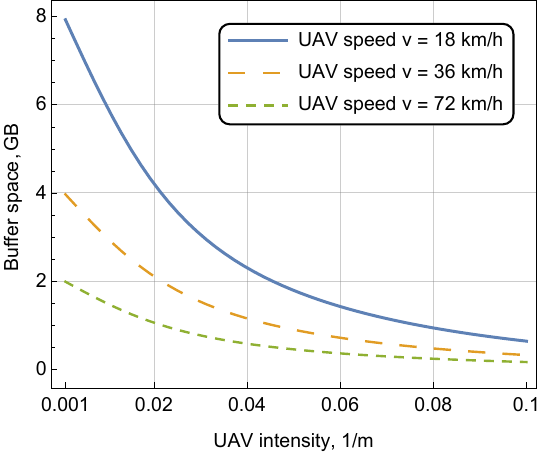}
		\label{fig:2_velocity_random}
	}\\
	\subfigure[{Controlled scheme}]
	{
		\includegraphics[width=0.75\columnwidth]{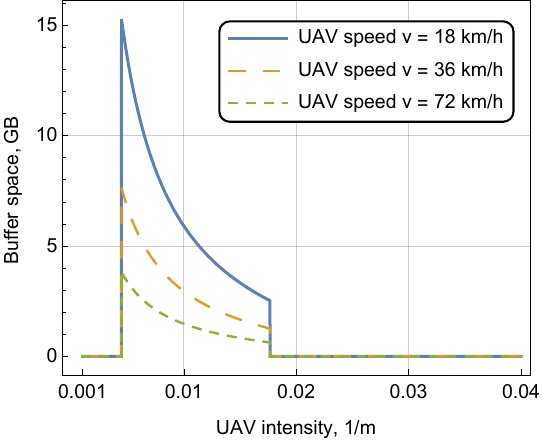}
		\label{fig:2_velocity_control}
	}
\caption{Impact of UAV velocity on required buffer space.}
\label{fig:2_velocity}
\end{figure}


We first assess the amount of buffer space required to temporarily store the data during a turn. Among the considered system parameters, velocity and deployment type are expected to primarily affect this buffer space. The impact of the UAV velocity, $v_U$, on the offloading factor of the routed and controlled schemes is displayed in Fig. \ref{fig:2_velocity} for $N=10$, $R=3$ Gbps, $4\times{}1$ transmit and receive antenna arrays, and urban dense deployment. Note that the buffer space requirements are independent of having route selection. Analyzing the data for the routed scheme, one may notice that higher UAV velocities drastically reduce the mean buffer space required for uninterrupted UAV `bridge' operation. Further, one may learn that buffering requirements also reduce significantly, when the UAV intensity increases. However, even for the worst considered case of $\lambda=0.001$ 1/m, the mean required buffer space is only 2 GB, which is an affordable value. Similar trends and performance figures are observed for the controlled scheme. The main difference is that with this scheme no buffering is needed when the UAV intensity reaches the value of $\lambda=1/2d$.

\begin{figure}[!t]
    \centering
    \includegraphics[width=0.75\columnwidth]{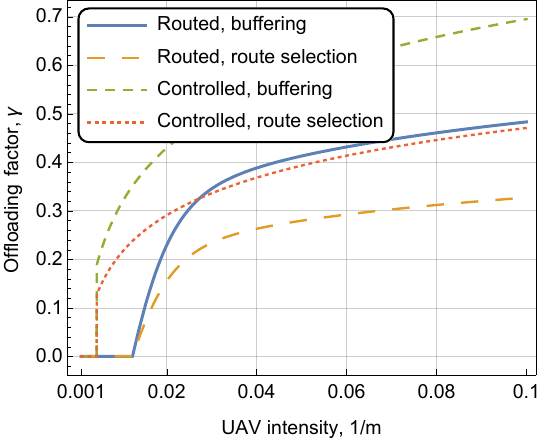}
    \caption{Offloading factor with and without buffering.}
    \label{fig:2_schemes}
    \vspace{-0mm}
\end{figure}


We now assess the gains of buffering at the UAVs. Recall that the use of buffering increases the UAV `bridge' availability by mitigating interruptions caused by LoS blockage. Theoretically, it implies that the offloading factor is now provided by $\gamma=\zeta\epsilon$ instead of $\gamma=\zeta\epsilon\nu$. To this end, Fig. \ref{fig:2_schemes} illustrates the offloading factor for the routed and controlled schemes with route selection and with/without buffering as a function of the UAV intensity, for $N=10$, $R=3$ Gbps, $h_U=30$ m, urban dense deployment type, and $4\times{}1$ transmit and receive antenna arrays.

Analyzing the presented data, one may notice that for both controlled and routed schemes buffering at the UAVs substantially improves the offloading gains. Particularly, for the considered system parameters, the routed scheme with buffering outperforms the controlled alternative without buffering starting from the UAV intensity of approximately $0.03$. Furthermore, the gap between the schemes with and without buffering increases as the UAV intensity grows. Hence, we may conclude that exploiting the full potential of UAV `bridging' by enabling buffering and route selection as well as by utilizing the controlled scheme, one may drastically improve the baseline system performance by offloading up to $30\%-70\%$ of traffic demand.

\subsection{Impact of Radio Parameters}


One of the benefits of the offloading schemes considered in this study is that NR modules utilized at the UAVs are conventional consumer-grade radios. However, it may be useful to understand the further benefits of employing more advanced NR equipment. We therefore proceed to study the effect of antenna arrays at the transmit and receive sides.

\begin{figure}[!t]
\centering	
	\subfigure[{Offloading factor}]
	{
    \includegraphics[width=0.75\columnwidth]{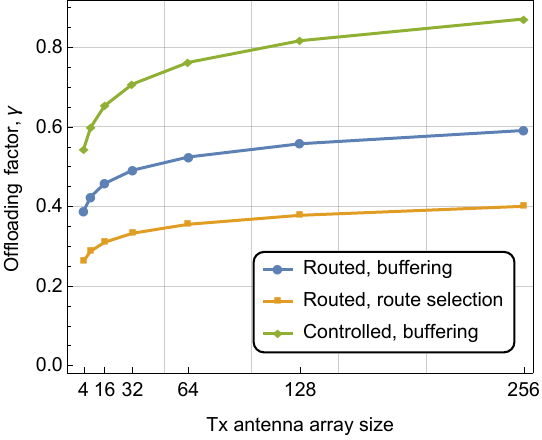}
    \label{fig:radio1}
	}\\
	\subfigure[{Buffer space}]
	{
		\includegraphics[width=0.75\columnwidth]{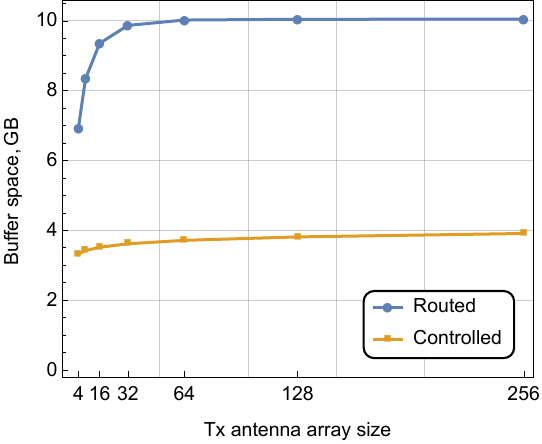}
		\label{fig:radio2}
	}
\caption{Impact of UAV antenna array.}
\label{fig:radio}
\end{figure}


The impact of the number of antenna elements forming the radiation pattern of the transmit antenna in the horizontal plane is highlighted in Fig. \ref{fig:radio1} for the routed and the controlled schemes with the above enhancements, $\lambda=0.04$~1/m, $N=10$, $R=3$~Gbps, $h_U=30$~m, and the urban dense deployment. As one may observe, increasing the number of antenna elements improves the offloading performance for all the considered schemes. Furthermore, the best gains in terms of the absolute numbers are observed for the controlled scheme with buffering. It is important to note that the gains are higher for the arrays having a smaller number of elements. Particularly, the use of $32\times{}1$ arrays instead of $4\times{}1$ ones improves the offloading factor from $0.55$ to $0.70$ for the controlled scheme with buffering and from $0.27$ to $0.34$ for the routed scheme with route selection. Similar improvements are observed for more sophisticated receive antenna arrays. However, analyzing Fig. \ref{fig:radio2}, we note that the offloading gains come at the expense of increased buffer space for both routed and controlled schemes. As one may notice, for the routed scheme the required buffer space plateaus at approximately $10$ GB, while the difference between the two schemes approaches $6$ GB.

\begin{figure*}[!t]
\centering	
	\subfigure[{Effect of block and street length}]
	{
		\includegraphics[width=0.31\textwidth]{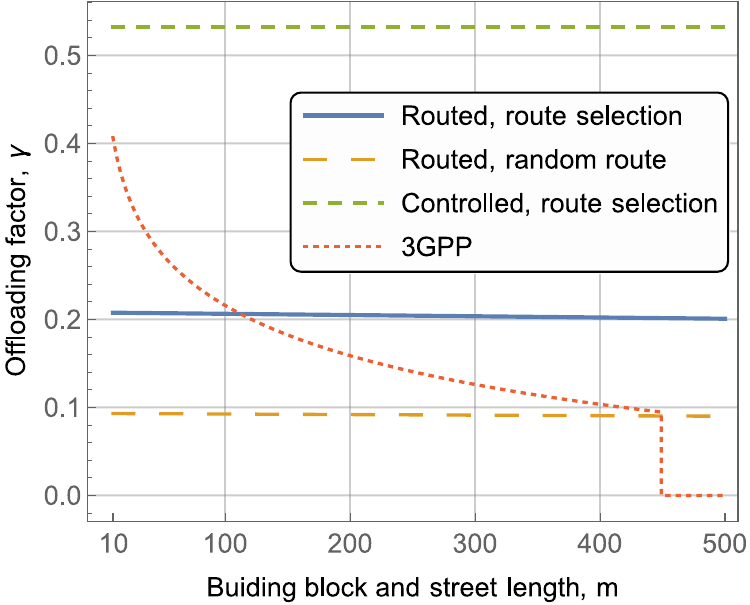}
		\label{fig:1_3gpp}
	}~
	\subfigure[{Effect of resources availability}]
	{
		\includegraphics[width=0.31\textwidth]{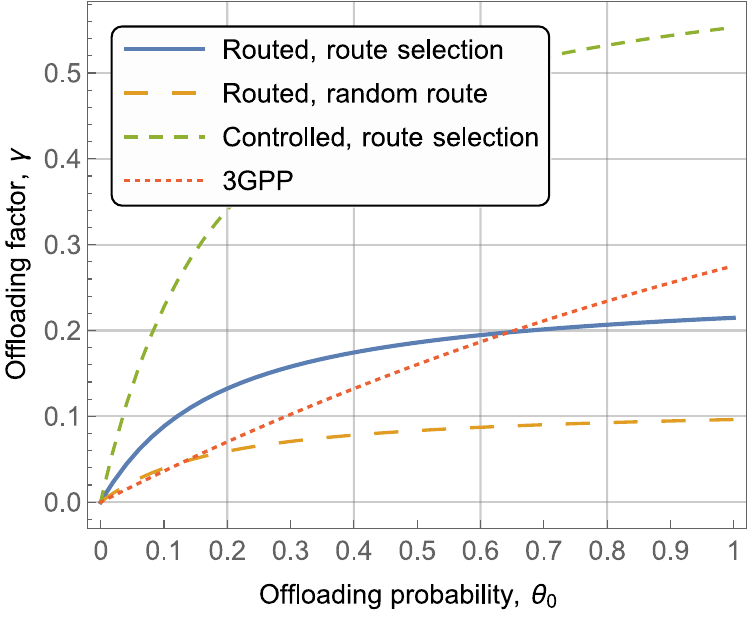}
		\label{fig:2_3gpp}
	}~
	\subfigure[{Effect of NR BS availability}]
	{
		\includegraphics[width=0.31\textwidth]{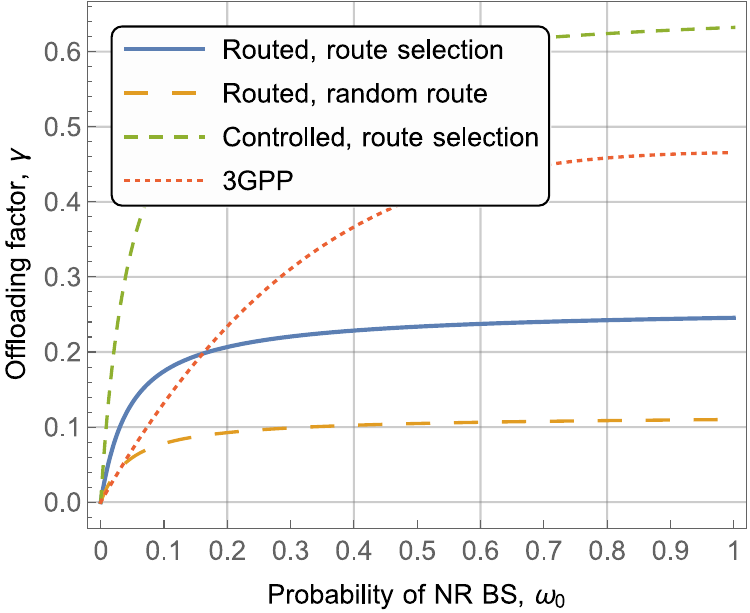}
		\label{fig:3_3gpp}
	}
\caption{Comparison between the proposed UAV `bridges' and standardized 3GPP IAB-based approaches.}
\label{fig:comparison}
\end{figure*}

\subsection{Comparison with 3GPP IAB-based Offloading}

To characterize the improvements brought by multi-hop UAV `bridges', we proceed by comparing the proposed method with the 3GPP-ratified IAB-based solution, where UAVs act as IAB nodes. Our introduced approach is a logical extension of the solution proposed by 3GPP for the case of multi-hop relaying, which is currently still under standardization. To perform this comparison, we modify the system model by assuming that an arbitrarily chosen NR BS has a sufficient amount of resources for traffic offloading with probability $\theta_0$. Furthermore, we also require that the NR BSs are deployed solely at the crossroads and the probability that there exists an NR BS at a crossroad is $w_0$. The offloading factor under these additional assumptions for our proposed and the 3GPP IAB-based options is derived in the Appendix.

The plots illustrating the performance of the proposed and the 3GPP IAB-based offloading schemes as functions of various parameters are provided in Fig. \ref{fig:comparison}. Analyzing the dependence of the offloading factor on the block and the street length shown in Fig. \ref{fig:1_3gpp} for $\lambda=0.001$ 1/m, one may observe that the performance of the 3GPP IAB-based approach is an exponentially decaying function. This behavior is explained by the propagation losses that increase with a growing distance from the source NR BS to the UAV IAB node and from the UAV IAB node to the destination NR BS. For shorter distances of up to $100$ m, it performs better as compared to the routed scheme. The reason is that at these distances the permanent availability of the IAB nodes impacts the offloading factor higher as compared to the increased distances between the involved entities. However, for the considered intensity of the UAVs in the area, the controlled scheme is significantly better than the 3GPP IAB-based solution across the entire range of the block and street lengths. This highlights the importance of aligning the distances between the UAVs in practical multihop UAV `bridging' implementations.

The results provided in Fig. \ref{fig:2_3gpp} indicate that the availability of the NR BS in the area having a sufficient amount of resources affects the proposed and the 3GPP IAB-based schemes differently. Particularly, the offloading factor for the 3GPP IAB-based solution improves approximately linearly with an increase in $\theta_0$, while the proposed scheme saturates by outperforming the 3GPP IAB-based approach at small values of $\theta_0$. This implies that the proposed option demonstrates better performance in overloaded conditions, which is of special importance for the network operators.

Finally, the data provided in Fig. \ref{fig:3_3gpp} indicates that the 3GPP IAB-based approach is more sensitive to the availability of the NR BSs in the environment as compared to the availability of radio resources at these NR BSs. In fact, the offloading factor for this scheme significantly exceeds that in the proposed routed solutions for all values of $w_0$ starting from $0.2$. However, the controlled scheme demonstrates significant gains by outperforming the 3GPP IAB-based approach by $0.1$ for $w_0=1$.

Summarizing these comparison results, we may conclude that for denser deployments and higher traffic conditions the controlled UAV `bridging' schemes significantly outperform the 3GPP IAB-based approaches. For sparse deployments, the routed schemes perform comparably to the 3GPP baseline. From the deployment point of view, the use of the 3GPP approach requires a fleet of UAVs provided by the network operator, which causes additional expenses on the fleet maintenance and incurs delays associated with the UAV deployment. On the other hand, the proposed approach is opportunistic in nature, as it depends on the availability of third-party UAVs within the area of interest. It reacts without delays and can be initiated when an overload is imminent. Therefore, the choice of the best solution depends not only on the system performance but also on the environmental and socioeconomic factors.

\section{Conclusions}\label{sect:conclusions}

In this work, we proposed a system design concept for multiple connected UAVs, termed the UAV `bridge'. We discussed the usage of UAV `bridges', which can offload excessive data traffic from an overloaded cell. Two types of schemes, namely, routed UAV `bridging' and controlled UAV `bridging' were considered for the data offloading procedure. Our developed methodology assessed the UAV `bridging' in terms of important system parameters and environmental characteristics, which include the deployment type, the requested data rate, the UAV features, and the number of antenna array elements.

We further tackled the connectivity properties of the multi-hop UAV `bridges' by capturing the fraction of time when the UAV `bridge' exists and the fraction of the offloaded data rate. By using the methods of stochastic geometry and renewal processes, we obtained the offloading gain of the multi-hop UAV `bridges' for both considered schemes. Our numerical results verified the performance of the offloading strategies for different input parameters. Moreover, we demonstrated that the use of buffering at the UAV side drastically improves the achievable offloading gains. 

\appendix

\subsection{Proposed UAV `Bridging' Approach} 

With several additional assumptions in mind, the offloading factor of our scheme can be written as (see subsection \ref{sect:framework}.A)
\begin{align}
\gamma=\zeta\epsilon\nu\theta,
\end{align}
where $\theta$ is the probability that at least one NR BS is having a sufficient amount of resources for traffic offloading.

Recalling our consideration that only those NR BSs that do not increase the flight time through the area of interest can be used for offloading, we now proceed by estimating $\theta$ as a function of $\theta_0$. First, we specify how many potential deployment locations (crossroads) can be used for offloading. Let $X$ and $Y$ denote the RVs specifying the location of the overloaded NR BS. These RVs have a uniform distribution over the N~$\times$~N grid of the considered area. The sought RV is immediately given by $(N-X)(N-Y)$. Observing that
\begin{align}
(N-X)(N-Y) = XY,
\end{align}
the pmf at hand can be formally written as
\begin{align}
u_i= \sum_{\forall k,l: k \cdot l=i} p_k p_l .
\end{align}

Note that the pmf of this distribution is difficult to estimate in practice. Therefore, we replace it with its continuous equivalent of the product of two uniform distributions. The CDF of such a distribution is readily given by
\begin{align}
F_Z(z) = \frac{z}{N^2} + \frac{z}{N^2} \log \frac{N^2}{z},
\end{align}
which leads to the following pdf
\begin{align}
f_Z(z) = \frac{1}{N^2} \log \frac{N^2}{z}.
\end{align}

The sought pmf of $(N-X)(N-Y)$ then reads as
\begin{align}
p_i = \int_{i+}^{i+1}f_Z(z)dz, i=1,2,\dots,N^2.
\end{align}

Now observe that at each potential deployment location, the probability of having an NR BS with a sufficient amount of resources is $w_0\theta_0$. Therefore, for any fixed number of potential deployment locations, the probability of having at least one of them with a sufficient amount of resources for traffic offloading is given by
\begin{align}
1-(1-w_0\theta_0)^{i},
\end{align}
which implies that the sought probability $\theta$ is
\begin{align}
\theta=\sum_{i=0}^{N^2}p_i[1-(1-w_0\theta_0)^{i}],
\end{align}
thus completing the analysis of our model.

\subsection{3GPP IAB-based Offloading}

Consider now the case of the standardized 3GPP offloading. Here, we assume that there are dedicated UAVs provided by the network operator to offload the traffic to the nearest NR BSs that have a sufficient amount of resources. Note that, as opposed to the proposed `bridge'-based offloading, there are no additional constraints on the choice of the NR BS, except for them being one-hop away from the target BS. In what follows, we consider the best-case scenario, where the overloaded NR BS is located in one of the inner grid points of the considered scenario, thus potentially having 4 NR BSs for traffic offloading and UAVs located midway between the overloaded NR BS and the target BS, i.e., at the distance of $(d_B+d)/2$. The latter assumption maximizes the throughput over the BS-UAV and UAV-BS links.

For the considered 3GPP-ratified offloading scheme, the offloading factor can be written as
\begin{align}
\gamma^{\star} = \zeta^{\star}\epsilon^{\star}\theta^{\star},
\end{align}
where $\zeta^{\star}$ is the probability that the link between the overloaded and the target NR BSs exists, $\epsilon^{\star}=\min\{1, R_{\max}/R\}$ is the fraction of the offloaded traffic, and $\theta^{\star}$ is the probability that there is at least one NR BS having a sufficient amount of resources for traffic offloading. As opposed to the `bridge'-based offloading, there is no factor accounting for the building blockage as there is always a LoS link between the overloaded NR BS and the UAV as well as the UAV and the target NR BS.

The probability $\theta^{\star}$ can be established similarly to the case of `bridge'-based offloading by taking into account the fact that there are only 4 NR BSs than can be used for offloading,~i.e.,
\begin{align}
\theta^{\star}=1-(1-w_0\theta_0)^{4}.
\end{align}

Further, note that $\zeta^{\star}$ is no longer an RV but rather a deterministic step function that depends on the distance $(d_B+d)/2$. When $(d_B+d)/2$ is greater than the distance that corresponds to an outage between the NR BS and the UAV, the link does not exist. Otherwise, it always exists. Finally, the offloading rate required to determine $\epsilon^{\star}$ can be obtained by using a half of the inter-BS distance, $(d_B+d)/2$. Combining these results, we obtain the offloading factor for the 3GPP-ratified solution.

\balance
\bibliographystyle{ieeetr}
\bibliography{multihop}


\begin{IEEEbiography}[{\includegraphics[width=1in,height=1.25in,clip,keepaspectratio]{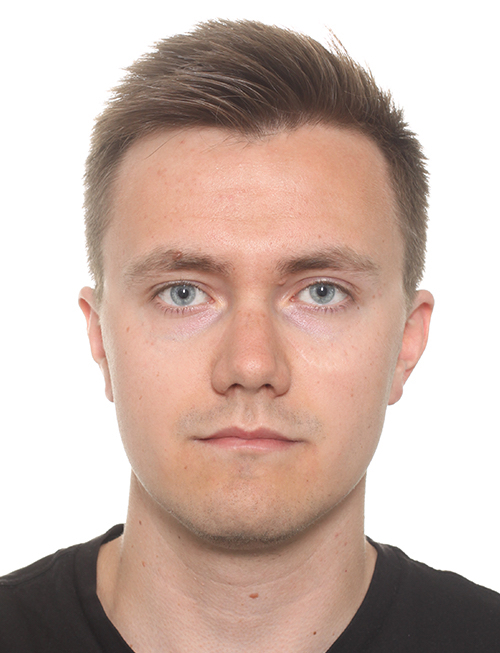}}]{Nikita Tafintsev} [S'20] (nikita.tafintsev@tuni.fi) is a Ph.D. candidate with the Unit of Electrical Engineering, Tampere University, Finland. He received M.Sc. degree with honors in Information Technology from Tampere University in 2019, and B.Sc. degree with honors in Radio Engineering, Electronics and Telecommunication Systems from Peter The Great St. Petersburg Polytechnic University, St. Petersburg, Russia, in 2017. His research interests include performance evaluation and optimization methods for 5G/5G+ wireless networks, applications of machine learning, and UAV communications.
\end{IEEEbiography}

\begin{IEEEbiography}[{\includegraphics[width=1in,height=1.25in,clip,keepaspectratio]{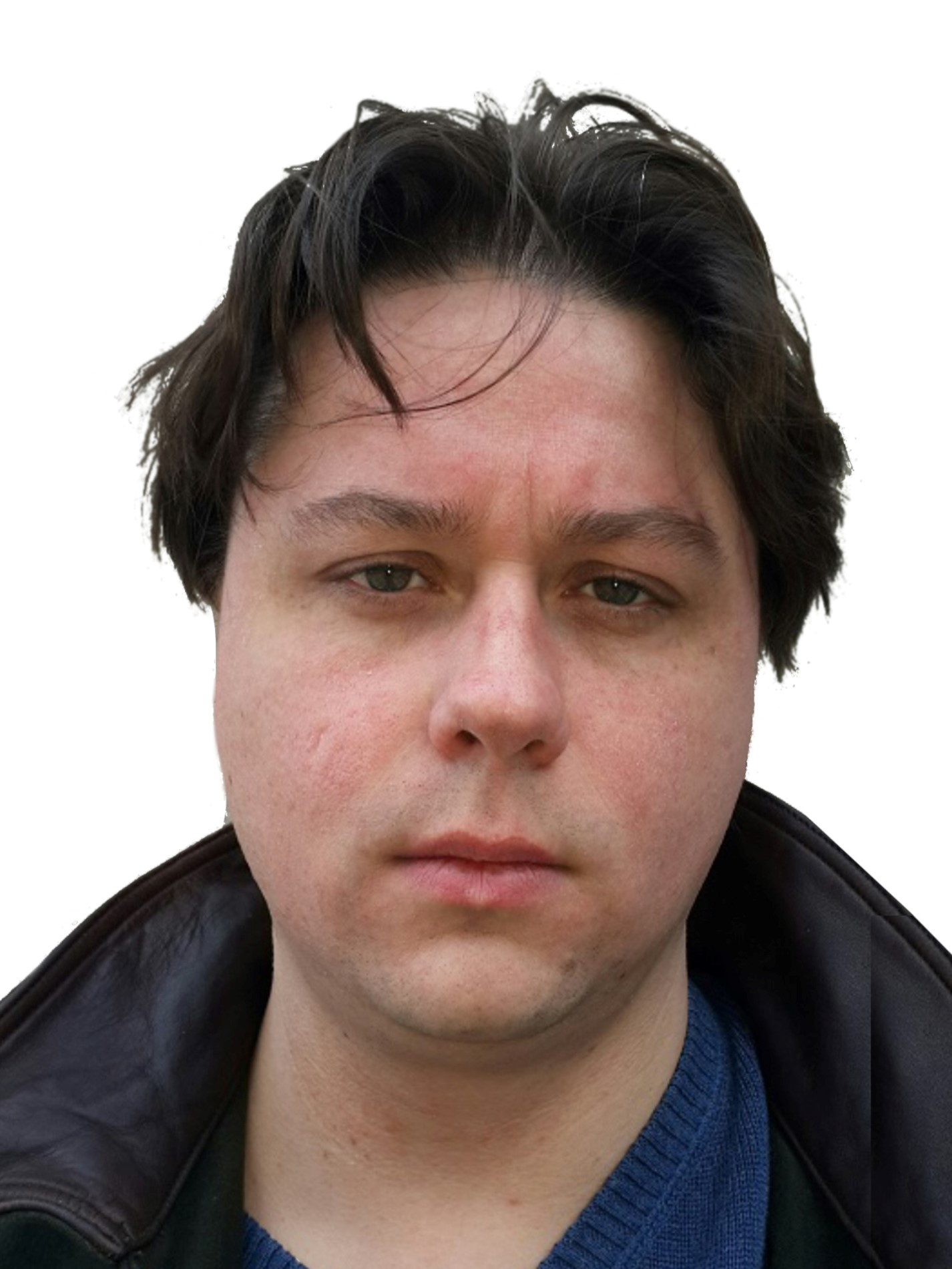}}]{Dmitri Moltchanov} (dmitri.moltchanov@tuni.fi) received the M.Sc. and Cand.Sc. degrees from St. Petersburg State University of Telecommunications, Russia, in 2000 and 2003, respectively, and the Ph.D. degree from Tampere University of Technology in 2006. Currently, he is a University Lecturer with the Faculty of Information Technology and Communication Sciences, Tampere University, Finland. He has (co-)authored over 150 publications. His current research interests include 5G/5G+ systems, ultra-reliable low-latency service, industrial IoT applications, mission-critical V2V/V2X systems, and blockchain technologies.
\end{IEEEbiography}

\begin{IEEEbiography}[{\includegraphics[width=1in,height=1.25in,clip,keepaspectratio]{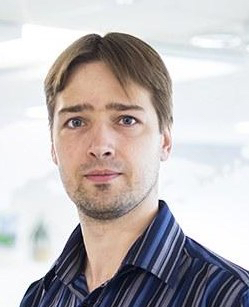}}]{Sergey Andreev} [SM'17] (sergey.andreev@tuni.fi) is an associate professor of communications engineering and Academy Research Fellow at Tampere University, Finland. He has been a Visiting Senior Research Fellow with King's College London, UK (2018-20) and a Visiting Postdoc with University of California, Los Angeles, US (2016-17). He received his Ph.D. (2012) from TUT as well as his Specialist (2006), Cand.Sc. (2009), and Dr.Habil. (2019) degrees from SUAI. He is lead series editor of the IoT Series (2018-) for IEEE Communications Magazine and served as editor for IEEE Wireless Communications Letters (2016-19). He (co-)authored more than 200 published research works on intelligent IoT, mobile communications, and heterogeneous networking.
\end{IEEEbiography}

\begin{IEEEbiography}[{\includegraphics[width=1in,height=1.25in,clip,keepaspectratio]{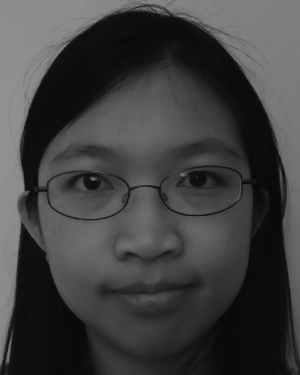}}]{Shu-ping Yeh} [M'06] (shu-ping.yeh@intel.com) received the B.S. degree from National Taiwan University in 2003, and the M.S. and Ph.D. degrees from Stanford University in 2005 and 2010, respectively, all in electrical engineering. She is currently a Research Scientist with Intel Labs. She specializes in wireless technology development for cellular and local area network access. Her recent research focus includes advanced self-interference cancellation technology, full-duplex PHY/MAC system designs, and multi-tier multi-RAT heterogeneous networks.
\end{IEEEbiography}

\begin{IEEEbiography}[{\includegraphics[width=1in,height=1.25in,clip,keepaspectratio]{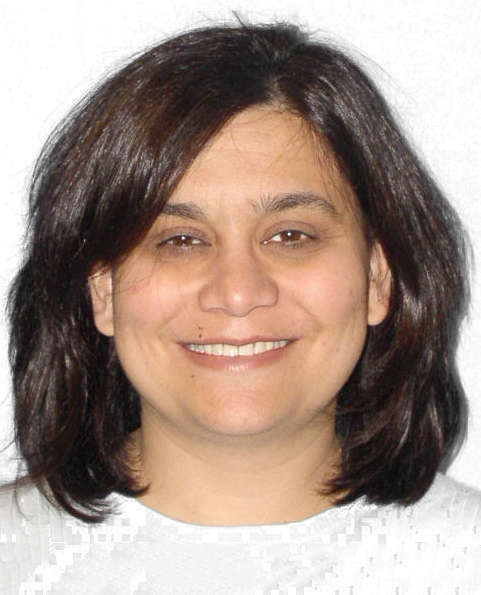}}]{Nageen Himayat} [M'88] (nageen.himayat@intel.com) received the B.S.E.E. degree from Rice University, the Ph.D. degree from the University of Pennsylvania, and the MBA degree from the Haas School of Business, University of California at Berkeley, Berkeley. She was with Lucent Technologies and General Instrument Corporation, where she developed standards and systems for both wireless and wireline broadband access networks. She is currently a Principal Engineer with Intel Labs, where she leads a team conducting research on several aspects of next generation (5G/5G+) of mobile broadband systems. She has authored over 250 technical publications, contributing to several IEEE peer-reviewed publications, 3GPP/IEEE standards, and numerous patents (28 granted, 55 pending). Her research contributions span areas such as multi-radio heterogeneous networks, mmWave communication, energy-efficient designs, cross-layer radio resource management, multi-antenna, and non-linear signal processing techniques.
\end{IEEEbiography}

\begin{IEEEbiography}[{\includegraphics[width=1in,height=1.25in,clip,keepaspectratio]{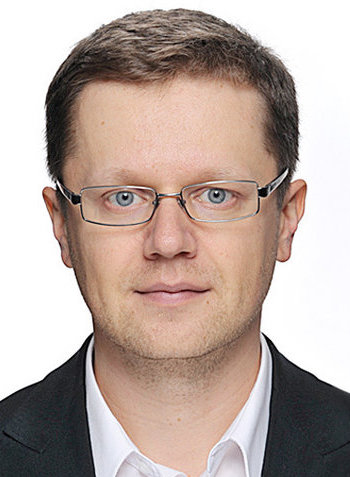}}]{Yevgeni Koucheryavy} [SM'09] (evgeny.kucheryavy@tuni.fi) received the Ph.D. degree from the Tampere University of Technology (TUT), Finland, in 2004. He is currently a Full Professor with the Unit of Electrical Engineering, Tampere University. He has authored numerous publications in the field of advanced wired and wireless networking and communications. His current research interests
include various aspects of heterogeneous wireless communication networks and systems, and emerging communication technologies for digitally augmented future-beings.
\end{IEEEbiography}

\begin{IEEEbiography}[{\includegraphics[width=1in,height=1.25in,clip,keepaspectratio]{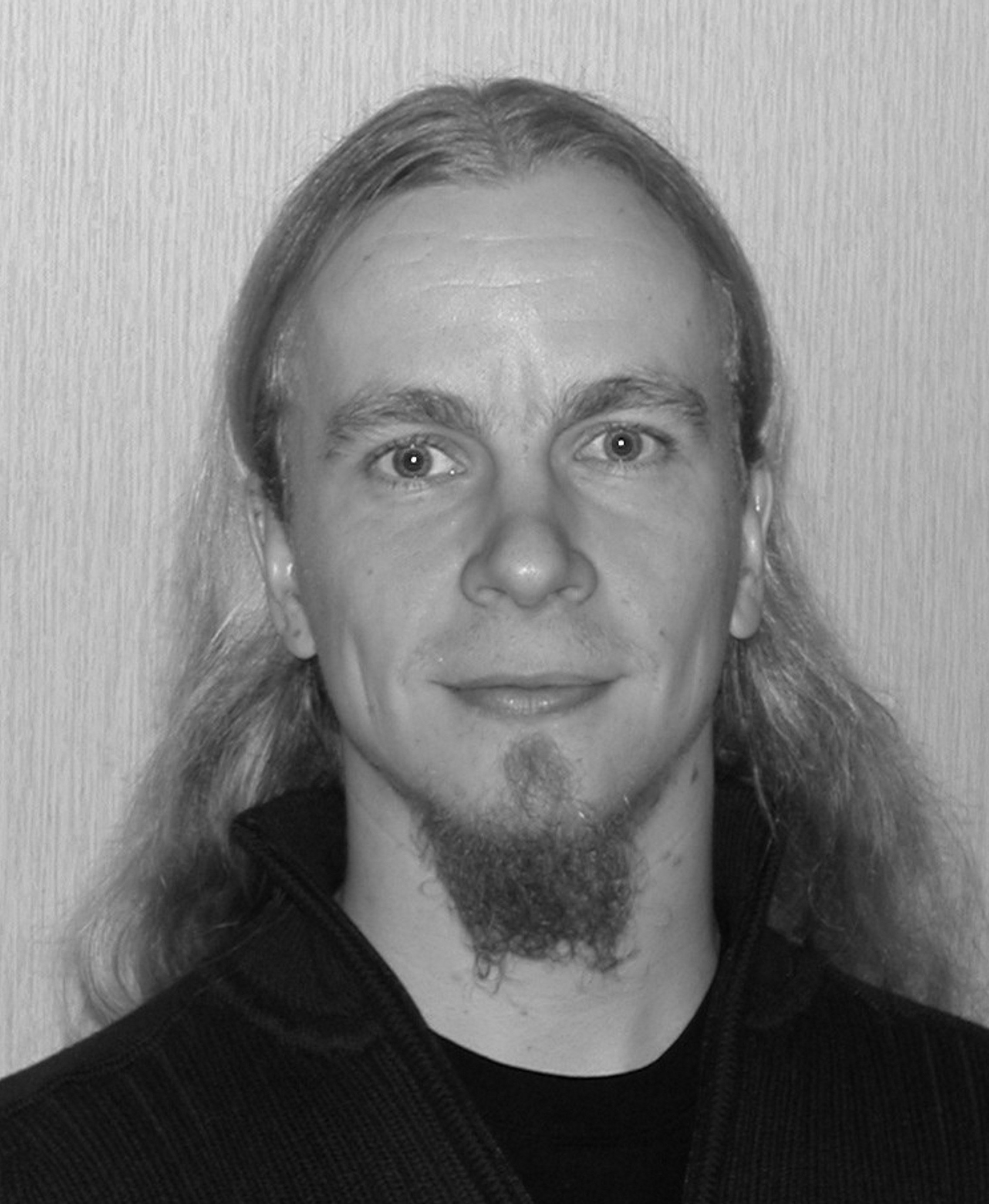}}]{Mikko Valkama} [SM'15] (mikko.valkama@tuni.fi) received the M.Sc. and D.Sc. degrees (Hons.) in electrical engineering (EE) from the Tampere University of Technology (TUT), Tampere, Finland, in 2000 and 2001, respectively. His Ph.D. dissertation was focused on advanced I/Q signal processing for wideband receivers: models and algorithms. In 2003, he was a Visiting Postdoctoral Research Fellow with the Communications Systems and Signal Processing Institute, San Diego State University (SDSU), San Diego, CA, USA. He is currently a Full Professor and the Unit Head of Electrical Engineering with Tampere University (TAU), Tampere. His current research interests include radio communications, radio localization, and radio-based sensing, with particular emphasis on 5G and beyond mobile radio networks. Dr. Valkama was a recipient of the Best Ph.D. Thesis Award of the Finnish Academy of Science and Letters for his Ph.D. dissertation.
\end{IEEEbiography}

\end{document}